\documentclass[a4paper,12pt]{article}
\pdfoutput=1 

\usepackage{jheppub} 

\usepackage[T1]{fontenc} 
\usepackage{feynmp}
\usepackage{slashed}
\usepackage{bbm}
\usepackage{graphicx}
\usepackage{hyperref}
\usepackage{cancel}
\usepackage{amssymb}
\usepackage{textcomp}
\usepackage{amsmath}
\usepackage{bm}
\usepackage{times}
\usepackage{epsfig}
\usepackage{color}

\title{ Observable Signatures of Scotogenic Dirac Model}

\author[1]{Shu-Yuan Guo}
\emailAdd{shyuanguo@pku.edu.cn}
\author[2]{Zhi-Long Han}
\emailAdd{sps\_hanzl@ujn.edu.cn}
\affiliation[1]{ Center for High Energy Physics, Peking University, Beijing 100871, China}
\affiliation[2]{ School of Physics and Technology, University of Jinan, Jinan, Shandong 250022, China}

\abstract{
In this work, we make a detailed discussion on the phenomenology of the scotogenic Dirac model, which could accommodate the Dirac neutrino mass and dark matter. We have studied the lepton-flavor-violating (LFV) processes in this model, which are mediated by the charged scalar $\phi^\pm$ and heavy Dirac fermions $N_i$. The experimental bounds, especially given by decays $\mu\to e\gamma$ and $\mu\to 3e$, have put severe constraints on the Yukawa couplings $y_\Phi$ and masses $m_{N1}$, $m_\phi$. We select the heavy Dirac fermion $N_1$ as dark matter candidate and find the correct relic density will be reached basically by annihilating through another Yukawa coupling $y_\chi$. After satisfying LFV and dark matter relic density constraints, we consider the indirect detections of dark matter annihilating into leptons. But the constraints are relatively loose, only the $\tau^+\tau^-$ channel can impose a mild excluding capability.  Then we make a detailed discussion on the dark matter direct detections. Although two Yukawa couplings can both contribute to the direct detection processes, more attention has been paid on the $y_\Phi$-related processes as the $y_\chi$-related process is bounded loosely. The current and future direct detection experiments have been used to set constraints on the Yukawa couplings and masses. The current direct detections bounds are relatively loose and can barely exclude more parameter region beyond the LFV. For the future direct detection experiments, the excluding capacities can be improved due to larger exposures. The detecting capabilities in the large mass region have not been weakened as the existence of mass enhancement from the magnetic dipole operator $\mathcal{O}_{\rm mag.}$. At last, we have briefly discussed the collider signal searching in this model, the most promising signature is pair produced $\phi^+\phi^-$ and decay into the signal of $\ell^+\ell^-+\slashed{E}_T$. The exclusion limits from collider on $m_{N1}$ and $m_\phi$ have provided a complementary detecting capability compared to the LFV and dark matter detections.}

\begin{document}
\maketitle
\flushbottom

\section{Introduction}
The success of neutrino oscillation experiments \cite{Fukuda:1998mi,Ahmad:2002jz} confirms the non-zero neutrino masses and mixings. Meanwhile, the nature of neutrinos, i.e., Majorana or Dirac, is still waiting  for certain positive signatures to confirm. Based on the $\Delta L=2$ Weinberg operator $\lambda LLHH/\Lambda$ \cite{Weinberg:1979sa}, the Majorana neutrino scenario attracts the most attention in early studies. The Weinberg operator can be realized by canonical seesaw \cite{Minkowski:1977sc,Mohapatra:1979ia,Schechter:1980gr,
Foot:1988aq}, low scale seesaw \cite{Mohapatra:1986aw,Mohapatra:1986bd,
Ma:2000cc,Malinsky:2005bi},  or radiative seesaw \cite{Zee:1985id,Babu:1988ki,
Krauss:2002px,Ma:2006km,Aoki:2008av,Cai:2017jrq}. The Majorana neutrino predicts the existence of lepton number violation (LNV) signatures, such as neutrinoless double beta decays ($0\nu\beta\beta$) \cite{Rodejohann:2011mu,Dolinski:2019nrj} and same-sign dilepton signature form heavy Majorana neutrino at collider \cite{Han:2006ip,Deppisch:2015qwa,Cai:2017mow}. However, the LNV signatures might be hard to detect. For instance, the effective neutrino mass relevant for $0\nu\beta\beta$ can vanish for normal neutrino mass ordering \cite{Pas:2015eia}, and the same-sign dilepton signature can also be suppressed by heavy Majorana neutrino beyond  TeV-scale \cite{Sirunyan:2018xiv}. An alternative approach to distinguish between the Majorana and Dirac nature of neutrinos is via the different capture rate $\Gamma(\nu_e+^3\!\text{H}\to ^3\!\text{He}+e^-)$ for cosmic neutrino background \cite{Long:2014zva,Zhang:2015wua,Roulet:2018fyh}.

Provided the neutrinos are Dirac particles, the neutrino masses in principle can be generated via direct Yukawa term $y_\nu\overline{F}_L H \nu_R$ with three copies of right-handed neutrinos $\nu_R$. Then in order to acquire sub-eV neutrino masses, the corresponding Yukawa coupling $y_\nu$ must be fine-tuned to $10^{-12}$ order. Hence, such direct Yukawa term is generally considered to be unnatural comparing with other SM fermions. Another issue is that $\nu_R$ is a neutral singlet under the SM gauge group, which means no symmetry can prevent the large Majorana mass term as in canonical seesaw mechanism \cite{Minkowski:1977sc,Mohapatra:1979ia}. Therefore, additional symmetry should be applied to protect the Dirac nature of neutrinos, such as $U(1)_{B-L}$ or $Z_N$ \cite{Ma:2016mwh}. Motivated by the above experimental and theoretical considerations, the Dirac scenario has been widely considered recently, and several models\cite{Gu:2006dc, Gu:2007ug, Farzan:2012sa, Chulia:2016ngi,
Bonilla:2016diq,Wang:2016lve,Borah:2017leo,Wang:2017mcy,CentellesChulia:2017koy,
Ma:2017kgb,Yao:2017vtm,Bonilla:2017ekt,Ibarra:2017tju,Borah:2017dmk,Das:2017ski,
Yao:2018ekp,CentellesChulia:2018gwr,CentellesChulia:2018bkz,Han:2018zcn,Borah:2018gjk,
Borah:2018nvu,Calle:2018ovc,Carvajal:2018ohk,Ma:2019yfo,Saad:2019bqf,Dasgupta:2019rmf,
Enomoto:2019mzl,Jana:2019mez,Ma:2019iwj,Ma:2019byo,
Restrepo:2019soi,CentellesChulia:2019xky,Calle:2019mxn} are proposed at tree or loop level.

Furthermore, the nature of dark matter (DM) is another open question in physics beyond SM. One promising candidate is the weakly interacting massive particles (WIMP) \cite{Jungman:1995df}. In the attractive scotogenic scenario, the WIMP DM emerges as a radiative neutrino mass messenger \cite{Ma:2006km}. Here, we will consider the scotogenic Dirac model proposed in Ref.~\cite{Farzan:2012sa}. Besides yielding  correct relic density, the DM is also expected to be observed by direct detection, indirect detection, and collider searches \cite{Arcadi:2017kky}. Since no concrete signature is observed yet, the direct detection experiments, e.g., PandaX-II \cite{Cui:2017nnn} and Xenon1T \cite{Aprile:2018dbl}, have set stringent bounds on the DM-nucleon scattering cross section. One appealing nature of the scotogenic model is that the DM-nucleon scattering processes are loop induced when considering fermion DM $N_1$ and thus with an extra natural $1/(16\pi^2)^2$ suppression of the cross section \cite{Schmidt:2012yg,Ibarra:2016dlb}. Even though suppressed, direct detection experiments are actually able to probe certain parameter space \cite{Herrero-Garcia:2018koq}.

The main purpose of this paper is to figure out the observable signatures  of the scotogenic Dirac model \cite{Farzan:2012sa} and derive corresponding parameter space. Due to different nature of fermion DM $N_1$ between scotogenic Dirac (Dirac $N_1$) \cite{Farzan:2012sa}  and scotogenic Majorana model (Majorana $N_1$) \cite{Ma:2006km}, the corresponding phenomenological predictions are expected to be distinguishable. For instance, the electromagnetic dipole moments vanish for Majorana $N_1$, hence it's natural to expect a larger DM-nucleon scattering cross section for a Dirac $N_1$. Meanwhile, the annihilation channel like $N_1\bar{N}_1\to \ell^+\ell^-$ is $s$-wave dominant for a Dirac $N_1$, thus could be further observed by indirect detection \cite{Bai:2014osa}. Therefore, the observation of Dirac fermion DM $N_1$ by ongoing experiments might provide us some indirect hints for the nature of Dirac neutrino.

The rest of the paper is organized as follows: an introduction to the scotogenic Dirac model and neutrino mass generation is given in Section \ref{sec:MD}; in Section \ref{sec:LFV}, we give a detailed discussion on the lepton-flavor-violating(LFV) processes in this model, and constraints from experiments have been given; the dark matter detections, i.e. annihilating to current relics, indirect and  direct detections, have been discussed in Section \ref{sec:DM}; finally, we present in Section \ref{sec:LHC} a brief discussion of LHC signature searching in this model; and we summarize our conclusions in Section \ref{sec:con}.

\section{Model Description}\label{sec:MD}

This model was originally proposed in Ref.~\cite{Farzan:2012sa}, to radiatively generate neutrino mass of Dirac type and at the same time provide dark matter candidate. Apart from the right-handed neutrino $\nu_R$, other particles introduced are vector-like fermions $N$ and two inert scalars $\chi$, $\Phi$. To let the one-loop mass appears as the lowest order, one need to forbid the tree-level mass term $\overline{\nu_L} H \nu_R$, with $H$ denoting the standard model Higgs doublet. The requirement of only Dirac masses exist would disallow all the Majorana masses of $\nu_L$, $\nu_R$ and $N$, one could own this to the existence of additional symmetries. The assignment of symmetries is actually not unique, in the original work, two kinds of $Z_2$ and one global $U(1)$ are imposed. In our case, we reduce it into $Z_3\times Z_2$, see Table.~\ref{tab:particles}. The assigning of $Z_3$ will forbid tree-level Dirac neutrino mass, and Majorana masses as well. While $Z_2$ will be used to stabilize the lightest $Z_2$ odd particles, which we choose $N_1$ as dark matter candidate.
\begin{table}[!h]
	\centering
	\caption{Particles in the scotogenic Dirac model.}
	\begin{tabular}{c|cc|cccc}
	\hline
							&	$F_L$				&	$H$					&	$\nu_R$				&	$N$						&	$\Phi$					&	$\chi$	\\	\hline\hline
	$SU(2)_L$		&	$2$					&	$2$					&	$1$						&	$1$						&	$2$						&	$1$		\\
	$U(1)_Y$			&	$-\frac{1}{2}$	&	$\frac{1}{2}$	&	$0$						&	$0$						&	$\frac{1}{2}$		&	$0$	\\ 	\hline
	$Z_3$				&	$0$					&	$0$					&	$\omega$		&	$\omega$	&	$\omega$		&	$0$		\\
	$Z_2$				&	$+$					&	$+$					&	$+$			&	$-$	&	$-$					&	$-$		\\	\hline
	\end{tabular}
	\label{tab:particles}
\end{table}

\unitlength=1mm
\begin{figure}
	\centering
	\includegraphics[width=0.45\textwidth]{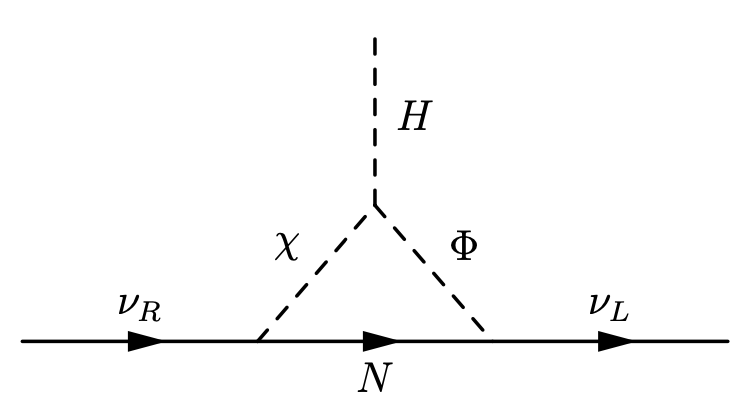}
\caption{One-loop neutrino mass in the scotogenic Dirac  model.}
\label{fig:neutrinomass}
\end{figure}

With the particles and symmetries given above, one could write down the relevant Yukawa Lagrangian as:
\begin{equation}
-\mathcal{L}_{\rm{Yuk.}} \supset \left(y_\Phi \overline{F_L} \tilde{\Phi} N + y_\chi \overline{\nu_R} N \chi + {\rm{h.c.}}\right)+ m_N\overline{N}N,
\end{equation}
with $\tilde{\Phi} \equiv i\sigma_2\Phi^\ast$. The general form of scalar potential is found to be:
\begin{eqnarray}
V &=& -\mu_H^2 H^\dagger H + \mu_\Phi^2 \Phi^\dagger \Phi + \frac{1}{2}\mu_\chi^2 \chi^2 + \frac{1}{2} \lambda_1 (H^\dagger H)^2 + \frac{1}{2} \lambda_2 (\Phi^\dagger \Phi)^2 + \frac{1}{4!} \lambda_3 \chi^4 \notag \\
& & + \lambda_4 (H^\dagger H)(\Phi^\dagger \Phi) + \frac{1}{2}\lambda_5 (H^\dagger H)\chi^2 + \frac{1}{2}\lambda_6 (\Phi^\dagger \Phi)\chi^2 + \lambda_7 (H^\dagger \Phi)(\Phi^\dagger H) \notag\\
& & +\left(\mu \Phi^\dagger H \chi + \rm{h.c.}\right).
\label{eq:potential}
\end{eqnarray}
For simplicity, we treat the singlet $\chi$ as real. The $\mu$-term will softly break the $Z_3$ symmetry, hence the parameter $\mu$ is natural to assume to be small. Generally, $\mu$ would be complex, but one has the degree of freedom to absorb its phase by a redefinition of $\Phi$ or $H$. Hence here we treat $\mu$ as real. After the spontaneous symmetry breaking, only $H$ develops out vacuum expectation value(VEV) $v$, the mass spectra are as follows\cite{Farzan:2012sa}:
\begin{eqnarray}
m_h^2 &=& 2\lambda_1 v^2,\\
m_{\phi^+}^2 &=& \mu_\Phi^2 + \lambda_4 v^2, \\
m_{\phi_I}^2 &=& \mu_\Phi^2 + \left(\lambda_4 + \lambda_7\right) v^2, \\
m_{(\chi, \phi_R)}^2 &=&
\begin{pmatrix}
\mu_\chi^2 + \lambda_5 v^2 & \sqrt{2}\mu v \\
\sqrt{2}\mu v & \mu_\Phi^2 + \left(\lambda_4 + \lambda_7\right) v^2
\end{pmatrix}.
\label{eq:mchiphiR}
\end{eqnarray}
Here $h$ labels the neutral real component of $H$, while $\phi_{R(I)}$ label the neutral real(imaginary) component of $\Phi$, the charged components of $\Phi$ are $\phi^\pm$. The $\lambda_7$ term would give a contribution to mass splitting between the neutral and charge components of $\Phi$. As we will not focus on this parameter region, we assume them to be degenerate, i.e. set $\lambda_7 = 0$. Since the charged scalar can be direct pair produced at colliders, the corresponding LEP bound, i.e., $m_{\phi^+}>80$ GeV \cite{Abbiendi:2013hk}, should be satisfied. ATLAS have searched a signature of $\ell^+\ell^-+\slashed{E}_T$ under the supersymmetry framework, which is similar to our case, and the mass of $\phi^\pm$ and $m_{N1}$ have been excluded till $550~\rm{GeV}$ and $300~\rm{GeV}$ \cite{Aad:2019vnb}, respectively. The $Z_2$ symmetry will forbid mixing between $H$ and $\Phi$, but mixing between $\Phi$ and $\chi$ is still exist, i.e. through the $\mu$ terms, with a mixing angle of
\begin{equation}
	\tan 2\theta = \frac{2\sqrt{2}\mu v}{\mu_\Phi^2 - \mu_\chi^2 + \left(\lambda_4 - \lambda_5\right) v^2}.
\end{equation}
As we commented above, the $\mu$-term is natural to be small, one would see below that another reason to ask for a small value of $\mu$ is from the tiny neutrino mass. The dark matter candidate could be, in principle, the lightest between $\chi$, $\phi_{R(I)}$, and $N_1$(the lightest $N$). For the case of $\chi$ or $\phi_{R(I)}$ to be dark matter candidate, Ref.~\cite{Gu:2007ug} had made a discussion. In such a case, the dark matter would mainly annihilate directly into SM Higgs or through the gauge boson. Bounds on dark matter mass and the relevant couplings were set by relic density and direct detection experiments \cite{Cline:2013gha,Arhrib:2013ela}. In this work, we consider an alternative case, i.e. select $N_1$ as the dark matter candidate.

The neutrino mass generation topology is depicted in Figure \ref{fig:neutrinomass}, and the mass is calculated as~\cite{Farzan:2012sa}:
\begin{eqnarray}
\left(M_\nu\right)_{\alpha\beta} = \frac{\sin2\theta}{32\pi^2\sqrt{2}}&&\sum_{k=1,2,3}(y_{\Phi})_{\alpha k} (y_{\chi}^\ast)_{\beta k}m_{N_k} \notag \\
&&\times\left(\frac{m_1^2}{m_1^2-m_{N_k}^2}\ln\frac{m_1^2}{m_{N_k}^2} - \frac{m_2^2}{m_2^2-m_{N_k}^2}\ln\frac{m_2^2}{m_{N_k}^2}\right),
\label{eq:numass}
\end{eqnarray}
$m_{1,2}$ here denote the mass eigenstates of the neutral scalars $\chi$ and $\phi_R$. The Yukawa $y_\Phi$ would couple $\Phi$ to SM lepton doublet, hence it is able to mediate the LFV processes. The stringent LFV experiment results would force it not to be large, which we would discuss later. Provided the dark matter $N_1$ accounts for the correct relic abundance, this would ask for a large Yukawa. Hence the annihilation may be mainly through the $y_\chi$ coupling. To arrive at the tiny neutrino mass, i.e. $m_\nu\sim 0.1~\rm {eV}$, the mixing between $\chi$ and $\Phi$ need to be pretty small, i.e. $\sin 2\theta \sim 10^{-7}$ for $y_\Phi\sim 0.1$, $y_\chi \sim 1$, and electroweak scale mass spectra. It's then proper to treat the matrix of Equation \ref{eq:mchiphiR} as diagonal. We label the two elements as $m_\chi^2$ and $m_\phi^2$, the latter will also denote the degenerate mass of various components of $\Phi$. To diagonalize the neutrino mass, two unitary matrices are needed, i.e. $m_\nu = U^\dagger M_\nu V$, here $U$ will rotate the left-handed $\nu_L$ into mass basis, it's just the well-known PMNS matrix, $V$ is the counterpart that rotates the right-handed neutrino $\nu_R$. The Yukawa couplings $y_\Phi$ and $y_\chi$ could be parametrized following a similar procedure as the Casas-Ibarra style\cite{Cerdeno:2006ha} in type-I seesaw:
\begin{equation}
y_\Phi = U D_{\sqrt{m_\nu}} R D_{\sqrt{M^{-1}}}, \qquad y_\chi = V D_{\sqrt{m_\nu}} S D_{\sqrt{M^{-1}}},
\label{eq:parametrization}
\end{equation}
the matrix $M$ is diagonal, which is defined as
\begin{equation}
M \equiv \frac{\sin2\theta}{32\pi^2\sqrt{2}} {\rm \ diag}(M_1,M_2,M_3),
\end{equation}
with
\begin{equation}
M_i = m_{N_i}\left(\frac{m_1^2}{m_1^2-m_{N_i}^2}\ln\frac{m_1^2}{m_{N_i}^2} - \frac{m_2^2}{m_2^2-m_{N_i}^2}\ln\frac{m_2^2}{m_{N_i}^2} \right).
\end{equation}
The two $D$ matrices in~\ref{eq:parametrization} indicate diagonal mass matrices with the diagonal entries being $\sqrt{m_\nu}$ and $\sqrt{M^{-1}}$, $R$ and $S$ are two arbitrary $3\times 3$ matrices and related to each other by $RS^\dagger = \mathbbm 1$.

\section{Lepton Flavor Violation}\label{sec:LFV}
The generation of neutrino mass relates closely to LFV processes. The neutrino oscillation between different flavors is one kind of flavor violating signature. Another signature is the flavor violating decay of charged leptons. Currently, experiments have been set up to probe various flavor violating decay of muon and tau lepton. No positive signal had been observed yet, hence turned into bounds on the decay branching ratios of those processes.

\begin{figure}
	\centering
	\includegraphics[width=0.45\textwidth]{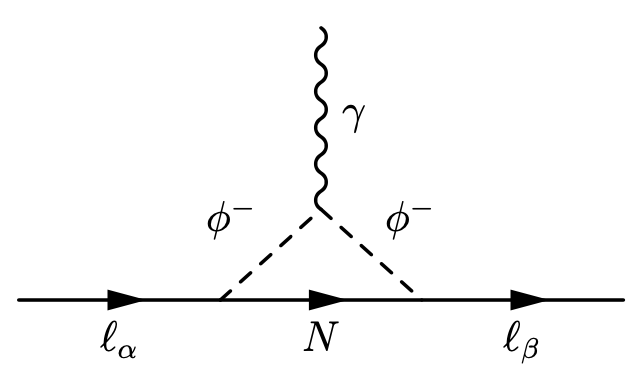}
	\caption{Diagram contributes to $\ell_\alpha\to\ell_\beta \gamma$.}
	\label{fig:mu2eg}
\end{figure}

The first kind of flavor violating process is the decay of $\ell_\alpha\to\ell_\beta \gamma$, $\alpha$ and $\beta$ stand for different flavors of charged leptons. The process is depicted in Figure~\ref{fig:mu2eg}. At the lowest order, only the Yukawa coupling $y_\Phi$ contributes to the decay, as $y_\chi$ couples to neutral $SU(2)_L$ singlets, which have no interaction with photon. From a view of effective field theory, such a process could be termed into an effective Lagrangian as~\cite{Kuno:1999jp}
\begin{equation}
\mathcal{L}_{\ell_\alpha\to \ell_\beta \gamma} = - \frac{4G_F}{\sqrt{2}} \left(m_\mu A_R \overline{\ell_{\beta L}} \sigma^{\mu\nu} \ell_{\alpha R} F_{\mu\nu} + {\rm h.c. } \right),
\end{equation}
with $A_R$ denoting the Wilson coefficient of the dipole operator. In this model, it's calculated as:
\begin{equation}
A_R = - \frac{\sqrt{2}}{8G_F} \frac{e (y_\Phi)_{\beta i} (y_\Phi^\ast)_{\alpha i}}{16\pi^2 m_{\phi}^2} j\left(\frac{m_{N_i}^2}{m_{\phi}^2}\right),
\end{equation}
where $j(r)$ labels the loop function:
\begin{equation}
j(r) = \frac{1-6r+3r^2+2r^3-6r^2\ln r}{12(1-r)^4}.
\end{equation}
The decay branching ratio is then given by
\begin{equation}
{\rm{Br}}(\ell_\alpha\to \ell_\beta \gamma)
= {\rm{Br}}(\ell_\alpha\to \ell_\beta \nu_\alpha \overline{\nu_\beta})\times \frac{3\alpha_{em}}{16\pi G_F^2} \left|\sum_i \frac{(y_{\Phi})_{\beta i} (y_{\Phi}^\ast)_{\alpha i}}{m_{\phi}^2} j\left(\frac{m_{N_i}^2}{m_{\phi}^2}\right)\right|^2.
\end{equation}
On the experimental side, the decay processes of $\mu\to e\gamma$, $\tau\to e\gamma$ and $\tau\to \mu\gamma$ have been probed, and upper limits on branching ratio are set as ${\rm{Br}(\mu\to e\gamma)}<4.2\times 10^{-13}$~\cite{TheMEG:2016wtm}, ${\rm{Br}(\tau\to e\gamma)}<3.3\times 10^{-8}$~\cite{Aubert:2009ag}, and ${\rm{Br}(\tau\to \mu \gamma)}<4.4\times 10^{-8}$~\cite{Aubert:2009ag}, respectively. The future detecting capabilities would be further improved to ${\rm{Br}(\mu\to e\gamma)}<6\times 10^{-14}$~\cite{Baldini:2013ke}, ${\rm{Br}(\tau\to e\gamma)}<3\times 10^{-9}$~\cite{Aushev:2010bq}, and ${\rm{Br}(\tau\to \mu \gamma)}<3\times 10^{-9}$~\cite{Aushev:2010bq}, respectively. The decay process contributes also to the muon anomalous magnetic moment, it's calculated as
\begin{equation}
\Delta a_\mu = m_\mu^2 \frac{4\sqrt{2}G_F}{e} {\rm Re} A_R = - \frac{(y_\Phi)_{ei} (y_\Phi^\ast)_{\mu i}}{16\pi^2 m_{\phi}^2} j\left(\frac{m_{N_i}^2}{m_{\phi}^2}\right),
\end{equation}
which contributes an opposite sign, hence can not be used to explain the muon anomalous magnetic moment.

\begin{figure}
	\centering
	\includegraphics[width=0.9\textwidth]{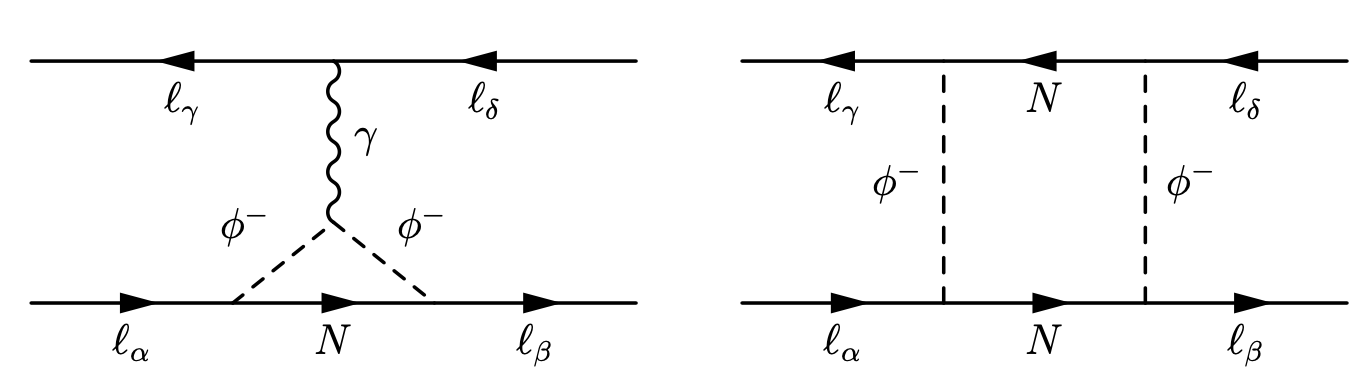}
	\caption{Diagrams contribute to $\ell_\alpha\to\ell_\beta\ell_\gamma\overline{\ell_\delta}$.}
	\label{fig:mu23e}
\end{figure}

\begin{figure}
	\centering
	\includegraphics[width=0.45\textwidth]{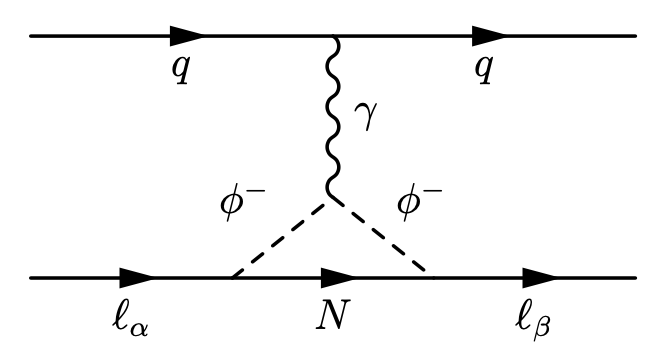}
\caption{Diagram contributes to coherent $\mu-e$ conversion in nuclei.}
\label{fig:mu2e}
\end{figure}

Another flavor violating decay process is $\ell_\alpha\to\ell_\beta\ell_\gamma\overline{\ell_\delta}$, two kinds of diagrams contribute to this decay in the model, see Figure~\ref{fig:mu23e} (contributions from the interchange between $\ell_\beta$ and $\ell_\gamma$ should also be taken into account when necessary). The most stringent constraint on the experiment comes from the decay of $\mu\to eee$, with an upper limit on the branching ratio as ${\rm{Br}}(\mu\to eee)<1.0\times 10^{-12}$~\cite{Bellgardt:1987du}, and the future detecting capability is expected to reach $\sim10^{-16}$~\cite{Blondel:2013ia}. Similar to the above case, one can write down an effective Lagrangian which describes the decay as~\cite{Kuno:1999jp}:
\begin{eqnarray}
\mathcal{L}_{\mu\to eee} &=& - \frac{4G_F}{\sqrt{2}} (m_\mu A_R \overline{e_L} \sigma^{\mu\nu} \mu_R F_{\mu\nu} + g_4 (\overline{e_L}\gamma^\mu \mu_L)(\overline{e_L}\gamma_\mu e_L)  \\\nonumber
&& \hspace{3em} + g_6 (\overline{e_L}\gamma^\mu \mu_L)(\overline{e_R}\gamma_\mu e_R) + {\rm h.c.}).
\end{eqnarray}
Apart from the dipole operator, the dim-$6$ vector operators can also contribute in, with the Wilson coefficients calculated in this model as
\begin{eqnarray}
g_4 &=& \frac{\sqrt{2}}{4G_F} \frac{1}{16\pi^2 m_{\phi}^2} \left[e^2 (y_\Phi)_{ei} (y_\Phi^\ast)_{\mu i} ~k\left(\frac{m_{N_i}^2}{m_{\phi}^2}\right)\right.
\\ \nonumber
&&\hspace{6em}\left.+(y_\Phi)_{ei} (y_\Phi^\ast)_{\mu i}(y_\Phi)_{ei} (y_\Phi^\ast)_{e i} ~l\left(\frac{m_{N_i}^2}{m_{\phi}^2}, \frac{m_{N_j}^2}{m_{\phi}^2}\right) \right], \qquad  \\
g_6 &=& \frac{\sqrt{2}}{4G_F} \frac{1}{16\pi^2 m_{\phi}^2} e^2 (y_\Phi)_{ei} (y_\Phi^\ast)_{\mu i} ~k\left(\frac{m_{N_i}^2}{m_{\phi}^2}\right).
\end{eqnarray}
The loop function $k(r)$ and $l(r_i,r_j)$ are read as
\begin{eqnarray}
k(r) &=& \frac{-2+9r-18r^2+11r^3-6r^3\ln r}{36(1-r)^4}, \\
l(r_i,r_j) &=& \frac{1}{4} \left(\frac{1}{(1-r_i)(1-r_j)} + \frac{r_i^2\ln r_i}{(r_i-r_j)(1-r_i)^2} - \frac{r_j^2 \ln r_j}{(r_i-r_j)(1-r_j)^2}\right).
\end{eqnarray}
Finally, the branching ratio for $\mu\to eee$ is given as~\cite{Kuno:1999jp}
\begin{eqnarray}
{\rm Br}(\mu \to eee) & = & \left[2 C_2 + C_4 + 32 \left(\ln\frac{m_\mu^2}{m_e^2} - \frac{11}{4}\right)C_5 + 16C_7 + 8C_9 \right] \\ \nonumber
&&~\times~{\rm Br}(\mu \to e\nu_\mu \overline{\nu_e}),
\end{eqnarray}
with the coefficients defined as
\begin{equation}
C_2 = |g_4|^2, C_4 = |g_6|^2, C_5 = |eA_R|^2, C_7={\rm Re}(eA_R^\ast g_4), C_9={\rm Re}(eA_R^\ast g_6).
\end{equation}

\begin{figure}
	\centering
	\includegraphics[width=0.45\textwidth]{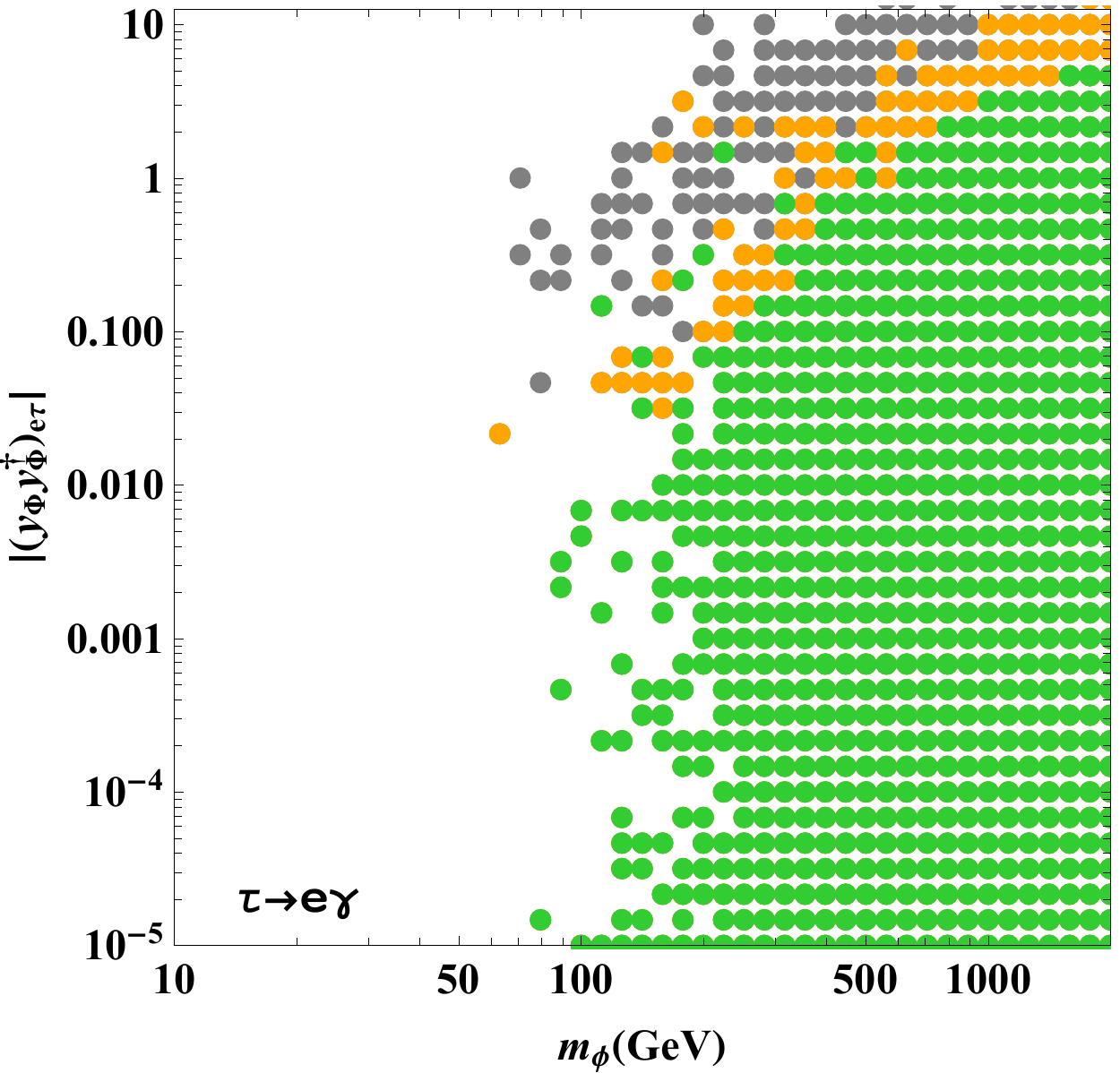}
	\includegraphics[width=0.45\textwidth]{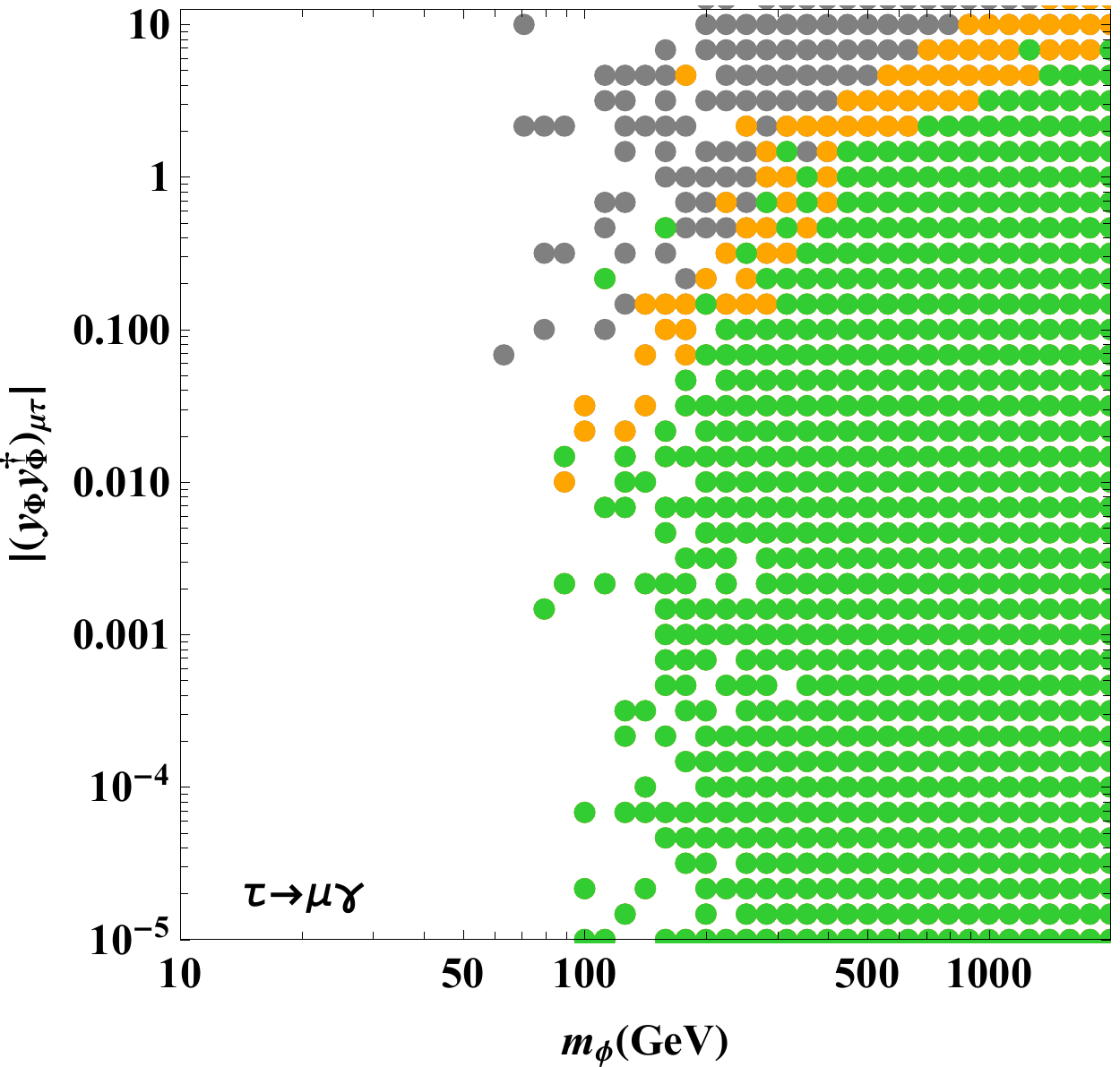}
	\includegraphics[width=0.45\textwidth]{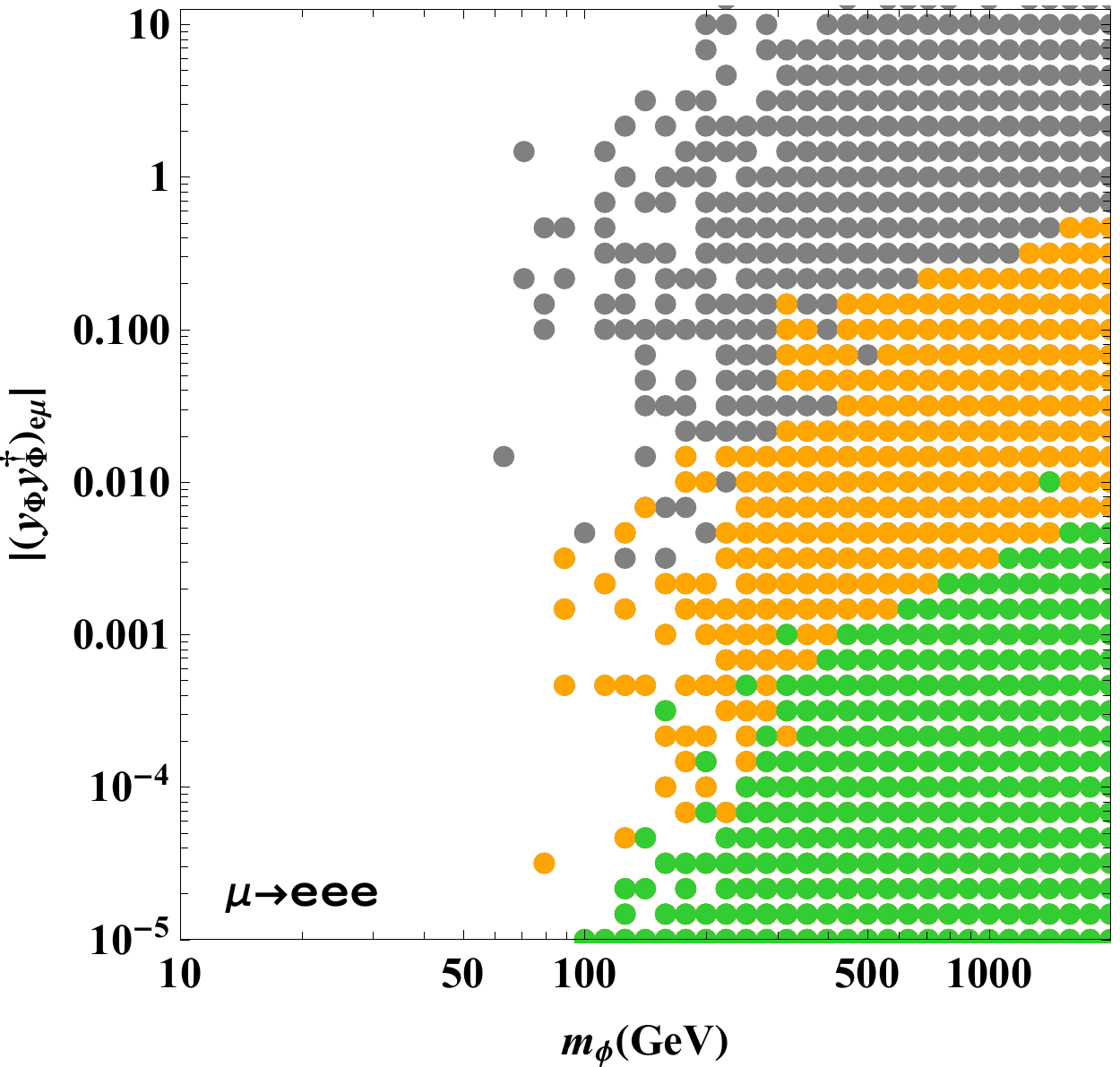}
	\includegraphics[width=0.45\textwidth]{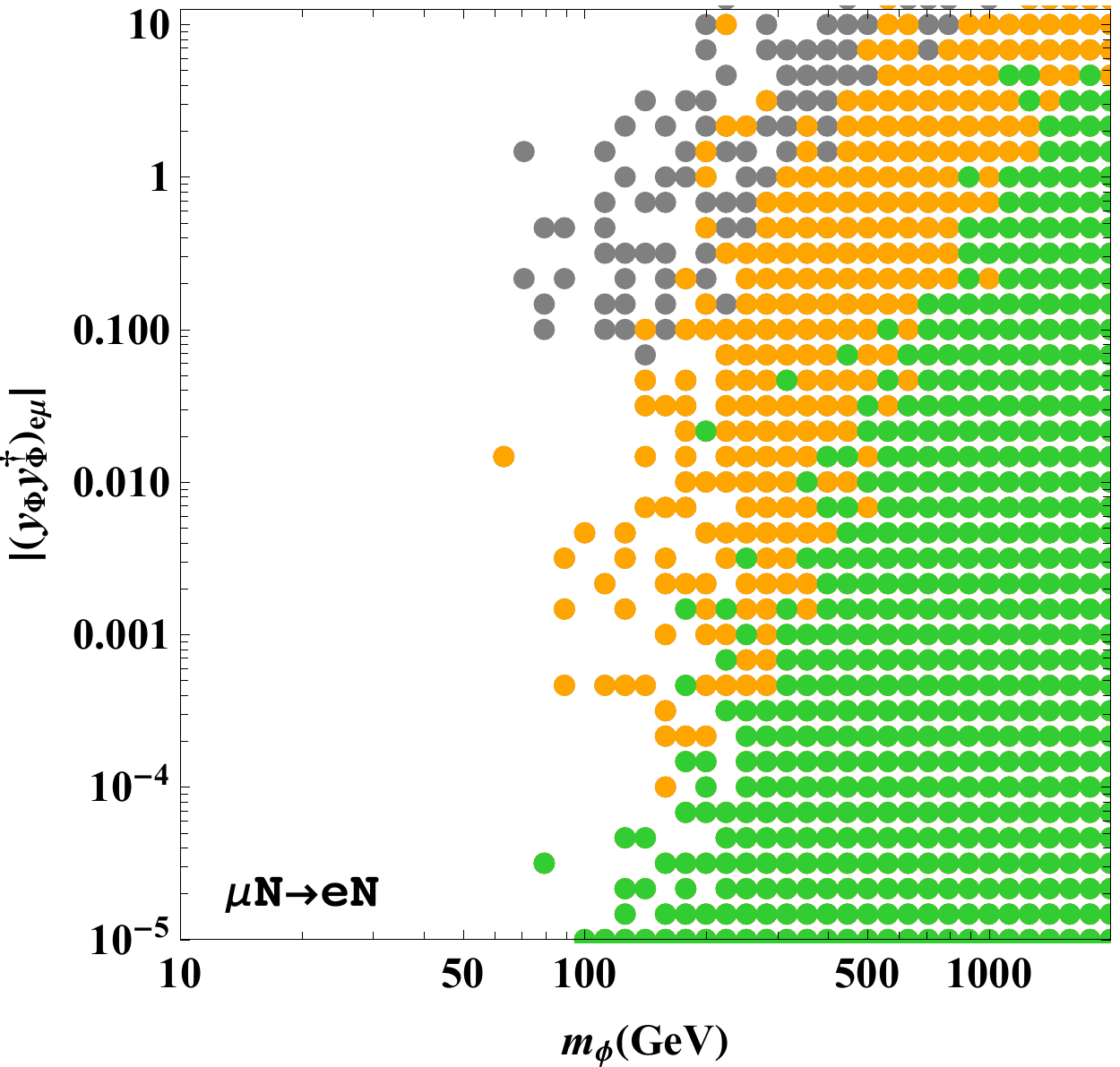}
	\includegraphics[width=0.45\textwidth]{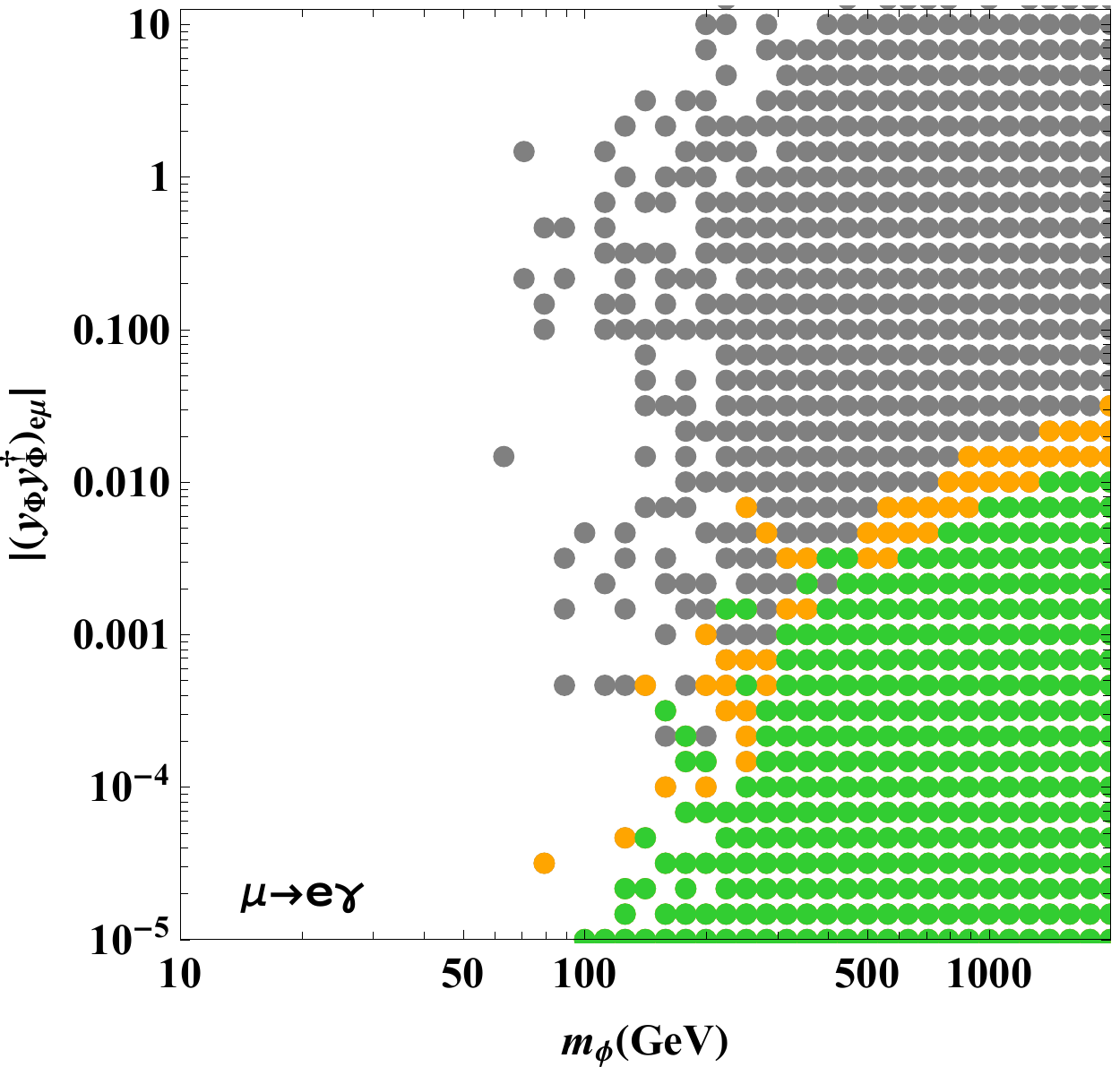}
	\includegraphics[width=0.45\textwidth]{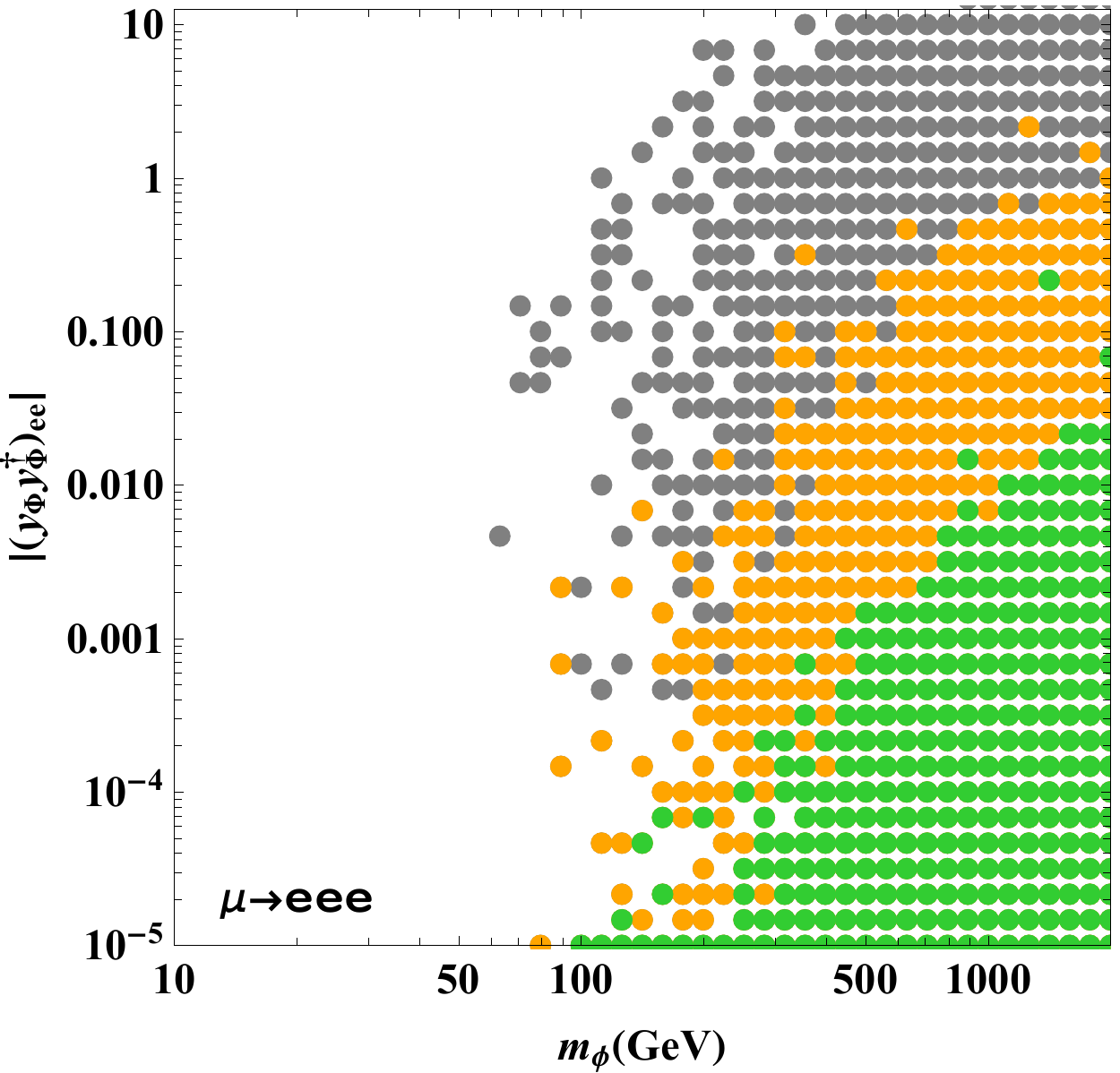}
\caption{The distributions of different entries of combination $y_\Phi y_\Phi^\dagger$. The yellow and green points are those that survived from current and future various experimental bounds, respectively. While the gray points have been excluded from current LFV experiments.}
\label{fig:LFV}
\end{figure}

\begin{figure}
	\centering
	\includegraphics[width=0.45\textwidth]{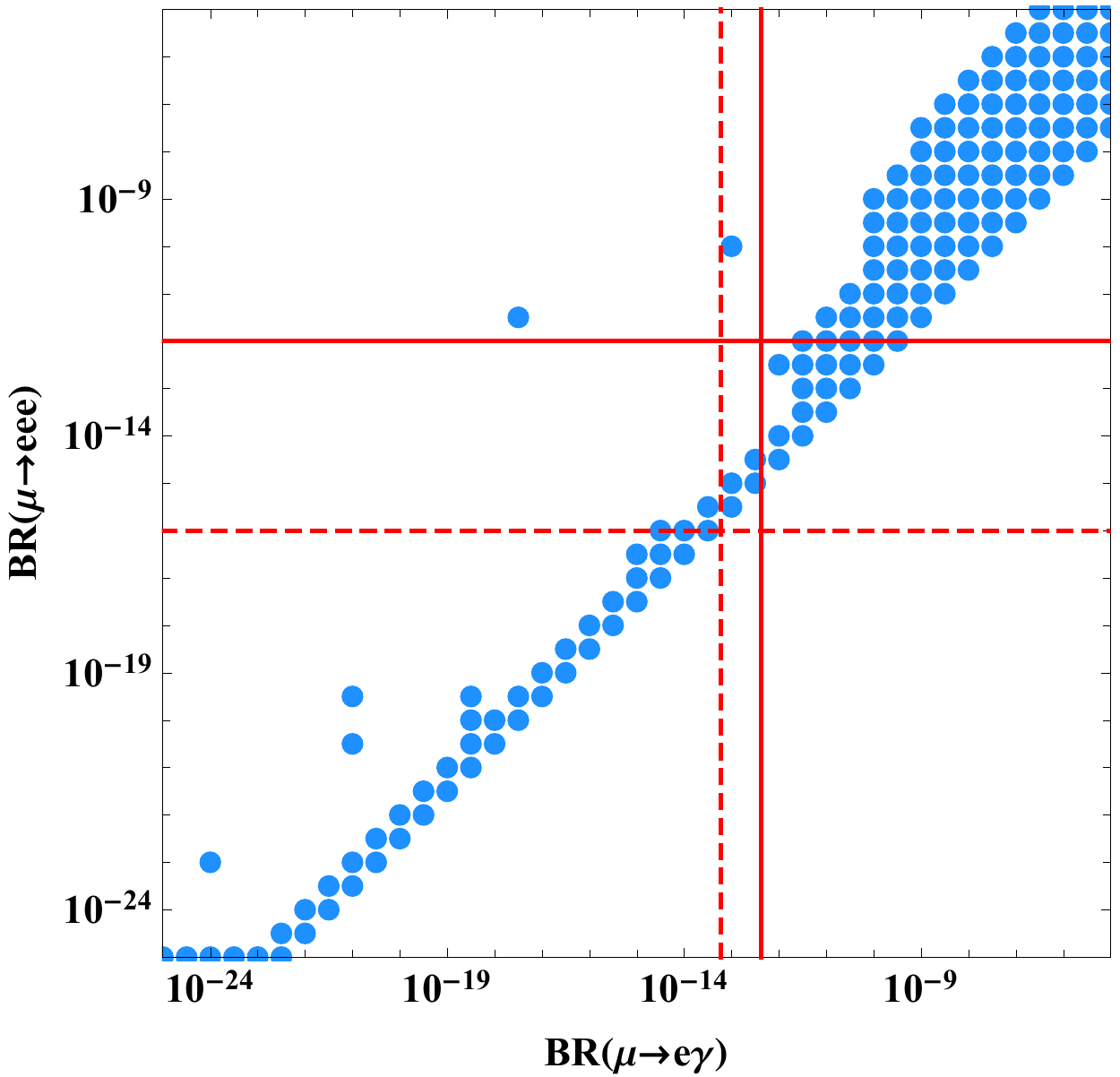}
	\includegraphics[width=0.45\textwidth]{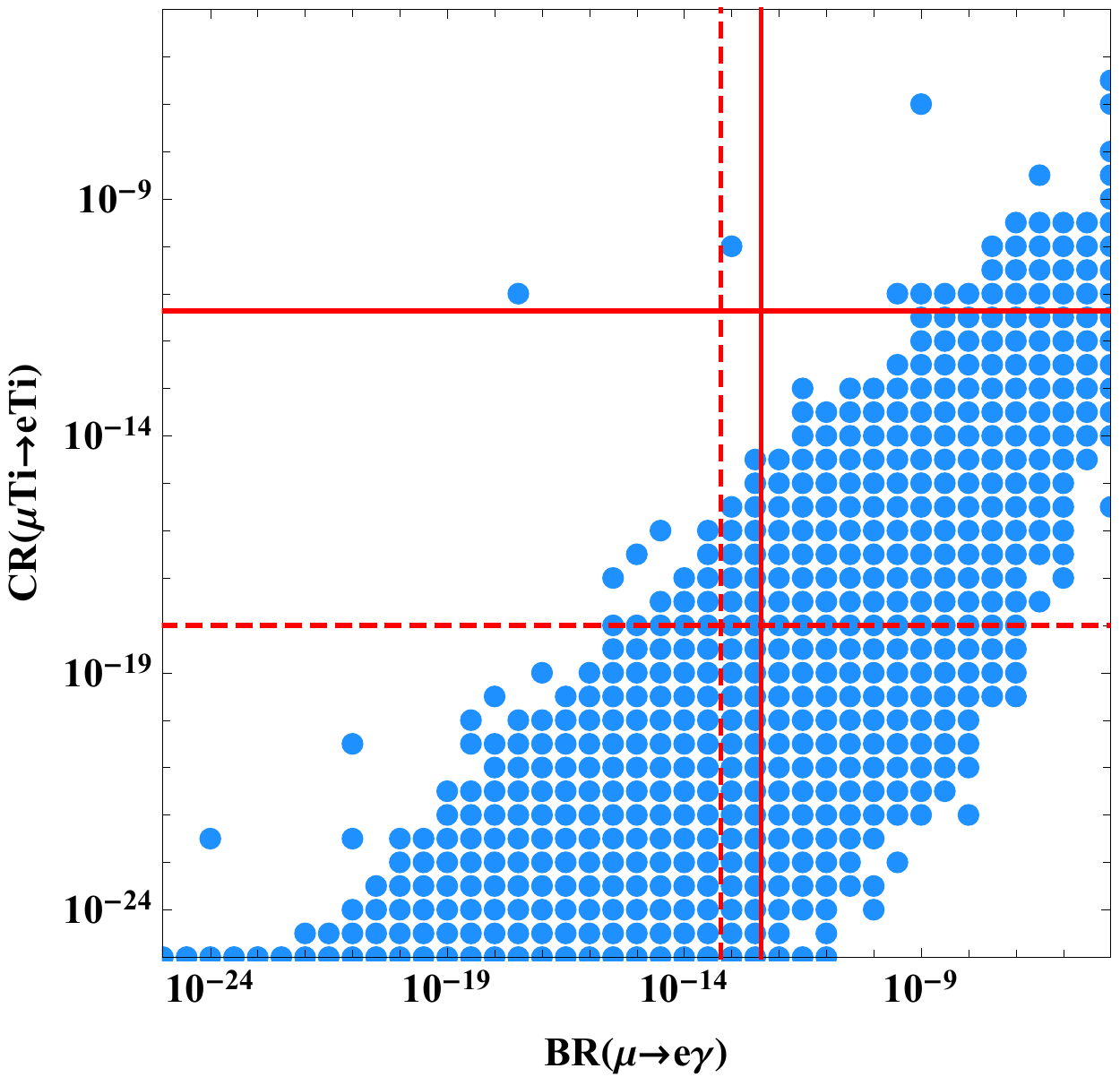}
	\caption{Distributions of branching ratio of the decays $\mu\to eee$, $\mu\to e\gamma$ and converting rate of process $\mu {\rm Ti} \to e {\rm Ti}$, before imposing the experimental constraints. The red solid lines stand for the corresponding bounds from current experiments, while the red dashed lines represent the future detection capabilities, respectively.}
	\label{fig:LFVbrs}
\end{figure}

The last bounded well process is the coherent $\mu-e$ conversion in nuclei. The converting diagram is depicted in Figure~\ref{fig:mu2e}, while the effective Lagrangian is given as~\cite{Kitano:2002mt}:
\begin{eqnarray}
\mathcal{L}_{\rm{conv}} &=& - \frac{4G_F}{\sqrt{2}} \left(m_\mu A_R \overline{e_L} \sigma^{\mu\nu} \mu_R F_{\mu\nu} + {\rm h.c.}\right) \\ \notag
&&- \frac{G_F}{\sqrt{2}}\sum_{q}\left[ g_{LV}^q (\overline{e_L}\gamma^\mu \mu_L)(\overline{q}\gamma_\mu q) + {\rm{h.c.}} \right].
\end{eqnarray}
The converting rate, relative to the muon capture rate, is then expressed as~\cite{Kitano:2002mt}
\begin{equation}
{\rm{CR}}(\mu N\to e N) = 2G_F^2 \lvert A_R D + \tilde{g}_{LV}^p V^p + \tilde{g}_{LV}^n V^n \rvert^2 \omega_{\rm {capt}}^{-1},
\end{equation}
with
\begin{equation}
\tilde{g}_{LV}^p =2 g_{LV}^u + g_{LV}^d = -4 g_6, \qquad \tilde{g}_{LV}^n = g_{LV}^u + 2 g_{LV}^d = 0,
\end{equation}
while the overlap integrals $D$, $V^p$, $V^n$, and muon capture rate $\omega_{\rm{capt}}$ could be found in Ref.~\cite{Kitano:2002mt}. The relative sign between $A_R$ and $\tilde{g}_{LV}^p$ is opposite, making a cancellation between these two terms. The converting had been probed on various isotopes, and bounds on converting rates are set as ${\rm CR}(\mu {\rm Ti} \to e {\rm Ti})<4.3\times 10^{-12}$~\cite{Dohmen:1993mp}, ${\rm CR}(\mu {\rm Au} \to e {\rm Au})<7.0\times 10^{-13}$~\cite{Bertl:2006up} and ${\rm CR}(\mu {\rm Pb} \to e {\rm Pb})<4.6\times 10^{-11}$~\cite{Honecker:1996zf}. The future detecting capability would be further improved to ${\rm CR}(\mu {\rm Ti} \to e {\rm Ti})< 10^{-18}$~\cite{Barlow:2011zza}, ${\rm CR}(\mu {\rm Al} \to e {\rm Al})<3\times10^{-17}$\cite{Litchfield:2014qea,Bartoszek:2014mya}.

We illustrate in Figure~\ref{fig:LFV} the constraints on the Yukawa couplings  from the above LFV processes. Various flavor violating processes have been marked explicitly in each subfigure. The points in gray are those excluded by the current experimental bounds of each process. Points in orange are in the detecting capabilities of future experiments while still allowed by the current constraints. The remaining green points are beyond the future detecting capabilities. In the numerical estimation, we have parametrized the Yukawa couplings as in Equation~\ref{eq:parametrization}, and use the latest, best fit, normal mass hierarchy, neutrino oscillation parameters from NuFIT group~\cite{Esteban:2018azc}:
\begin{eqnarray}
&&\theta_{12} = 33.82^\circ, \quad \theta_{23} = 48.6^\circ, \quad \theta_{13} = 8.60^\circ, \quad \delta_{\rm CP}= 221^\circ, \notag\\
&&\delta m^2 = 7.39\times 10^{-5}\ {\rm eV^2}, \quad \Delta m^2 = 2.528\times 10^{-3}\ {\rm eV^2}.
\end{eqnarray}
The other masses are set as
\begin{eqnarray}
&& 10^{-6}~{\rm eV}<m_{\nu 1} < 0.1~{\rm eV}, \quad 10\ {\rm{GeV}} < m_{N1}<1000\ {\rm GeV},\notag\\
&&m_{N1} < m_{\chi} < 2000\ {\rm GeV}, \quad  m_{N1} < m_{\phi} < 2000\ {\rm GeV}, \notag \\
&&m_{N1} < m_{N2}  < m_{N3}<2000\ {\rm GeV},
\label{eq:massset}
\end{eqnarray}
here $m_{\nu 1}$ stands for the lightest neutrino mass, and the summation of neutrino masses is constrained by the cosmological limit $\sum m_{\nu i}<0.12~\rm{eV}$ \cite{Aghanim:2018eyx}, the mixing angle between $\chi$ and $\phi_R$ is set as $\sin 2\theta \in \left[10^{-10}, 10^{-5}\right]$. For the two matrices $S$ and $R$, the real and imaginary parts of $S_{ij}$ are randomly and logarithmically taken values in $[10^{-3},10^3]$ and $R$ is acquired by the relation $RS^\dagger=\mathbbm 1$. For different choices of $m_\chi$ and $m_\phi$, some may introduce a negative mass in Equation \ref{eq:numass}. It can keep positive-definite through redefinition of neutrino field, i.e. $\nu_R\to -\nu_R$. We have checked that the redefinition would not have influences on our results.

From Figure~\ref{fig:LFV}, one could see the most stringent bounds are from $\mu\to e\gamma$ and $\mu\to eee$ decays. The $\mu\to e\gamma$ decay is related to $(y_\Phi y_\Phi^\dagger)_{e\mu}$, which has been bounded to be smaller than $\sim 0.01$ for $m_\phi<1500~\rm GeV$. The $\mu\to eee$ decay is related to both $(y_\Phi y_\Phi^\dagger)_{ee}$ and $(y_\Phi y_\Phi^\dagger)_{e\mu}$, with a big part of the parameter region going to be excluded. Decays of tau lepton are bounded loosely, which is obviously shown in the first row in Figure~\ref{fig:LFV}. For the coherent $\mu-e$ conversion process, we have mentioned above that there exist cancellation between the $A_R$ and $\tilde{g}_{LV}^p$ terms. From the fourth sub-figure we could see that the cancellation have obviously weakened constraints on $(y_\Phi y_\Phi^\dagger)_{e\mu}$.

\begin{figure}
	\includegraphics[width=0.45\textwidth]{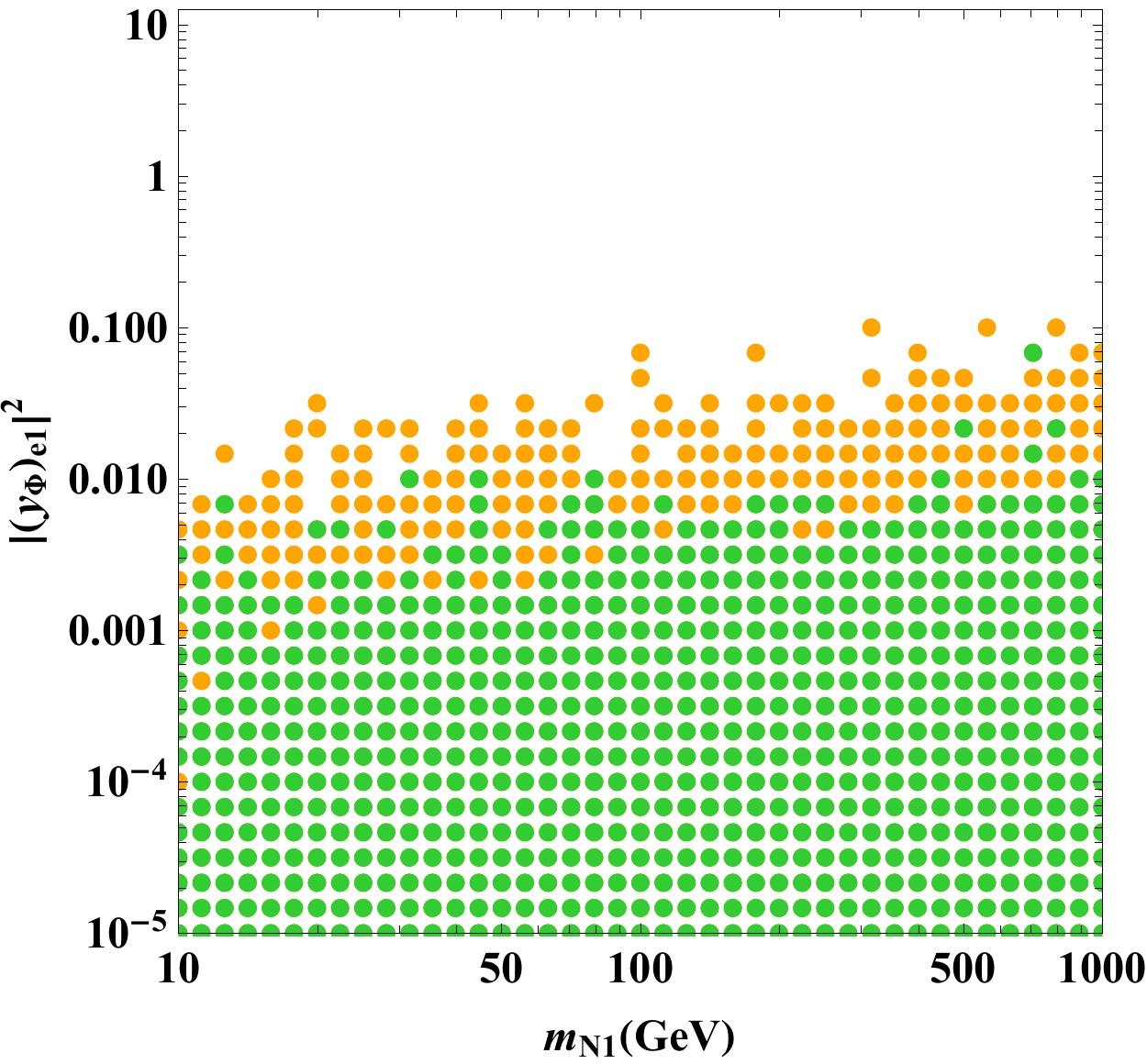}
	\includegraphics[width=0.45\textwidth]{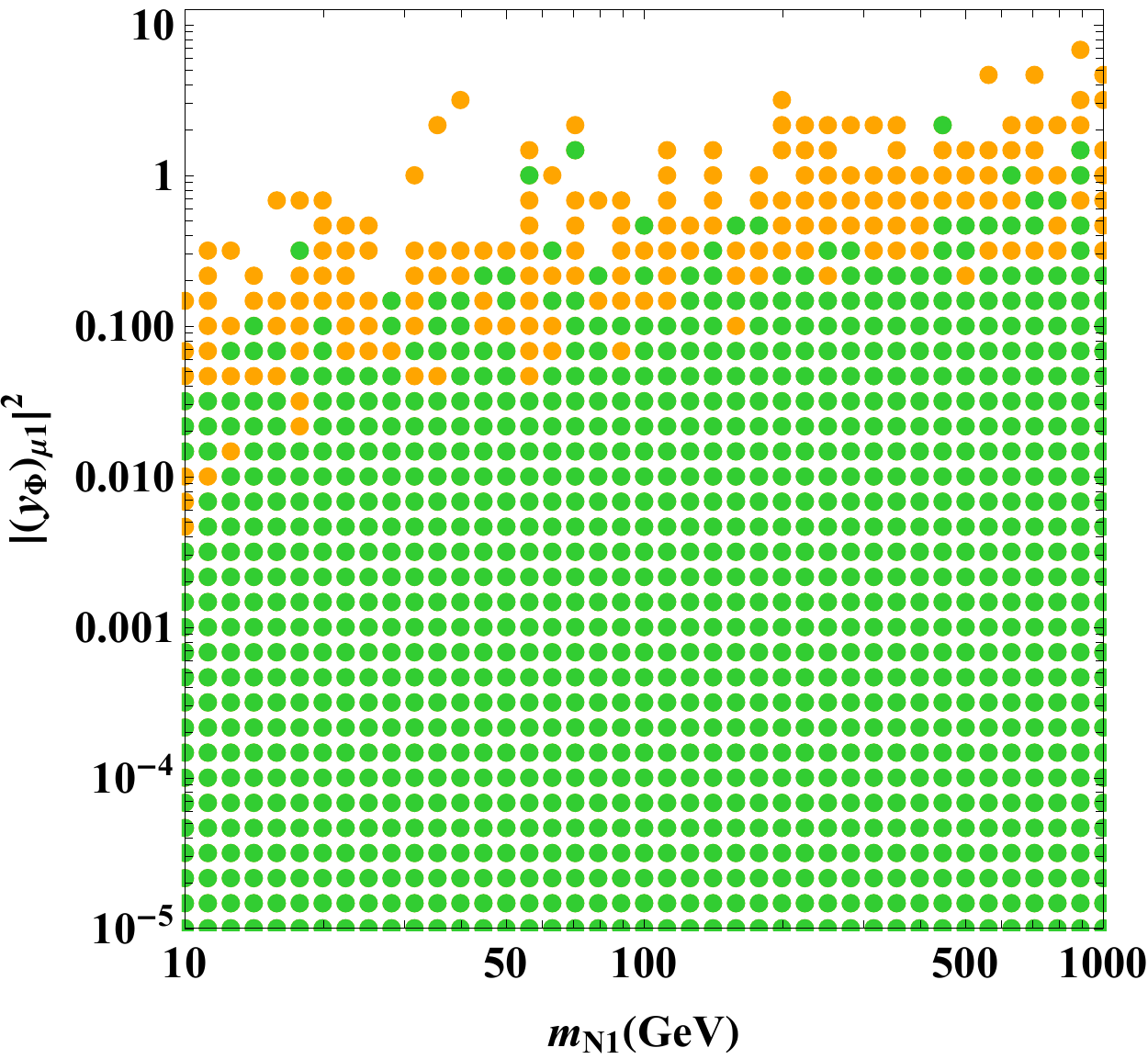}
	\includegraphics[width=0.45\textwidth]{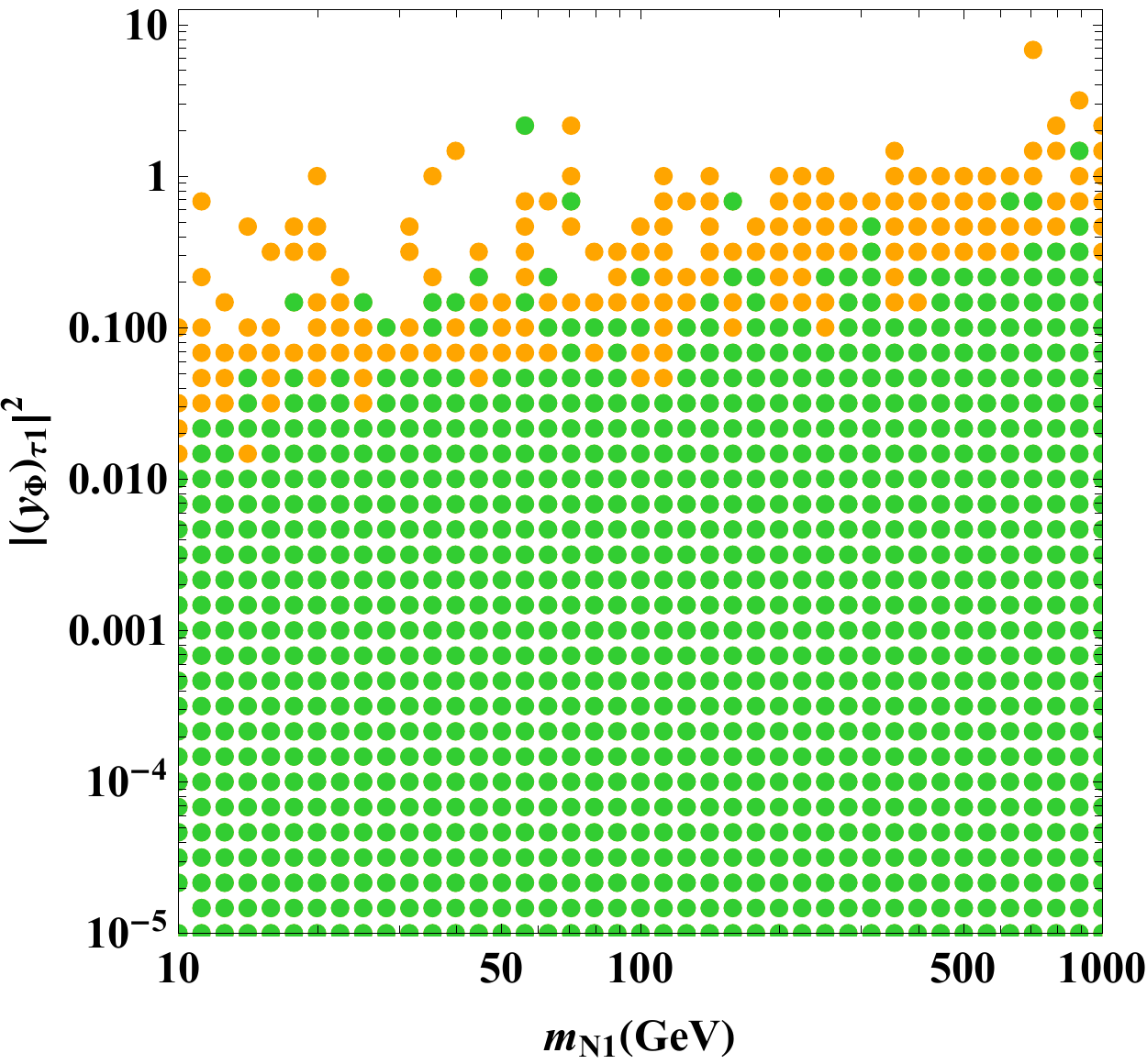}
\caption{The distributions of $\lvert y_\Phi \rvert_{\alpha 1}^2$ after satisfying all the LFV constraints.}
\label{fig:yyi1LFV}
\end{figure}

To have a clearer view on the excluding capabilities of various LFV processes, we present in Figure \ref{fig:LFVbrs} the distributions of ${\rm Br}(\mu \to eee)$, ${\rm Br}(\mu \to e\gamma)$ and ${\rm CR}(\mu {\rm Ti} \to e {\rm Ti})$. The red solid lines stand for the current experiment bounds on each process, while the red dashed lines represent detecting capabilities from the future experiments. One could see that for the current experiment bounds, decays $\mu\to e\gamma$ and $\mu \to eee$ are much tighter than the $\mu-e$ conversion process, and $\mu \to e\gamma$ is even tighter than $\mu \to eee$. For the future experiments, comparable detecting capabilities are provided by $\mu\to e \gamma$ and $\mu \to eee$, the $\mu-e$ conversion process could also show a good excluding ability, but the cancellation we mentioned above makes its bound still weaker than the decay $\mu\to e\gamma$. One could see that both ${\rm Br}(\mu \to e\gamma)$ and ${\rm CR}(\mu {\rm Ti} \to e {\rm Ti})$ increased as ${\rm Br}(\mu \to eee)$ increased, this is because they all have the contribution from triangle diagrams. The decay $\mu\to eee$ has an additional contribution from box diagrams, but they will become subdominant when Yukawa couplings are small, as one could confirm that an approximated linear relation becomes more obvious at the small branching ratios region, such behavior is consistent with the result in Ref.~\cite{Toma:2013zsa}.

Since $(y_\phi)_{\ell 1}$ is involved in DM $N_1$ annihilation, we give the element squares of the first column $\lvert (y_\Phi)_{\ell 1}\rvert^2$ in Figure~\ref{fig:yyi1LFV}  for illustration, after imposing all the current and future LFV constraints. One could see that a hierarchical distribution exists, the $\lvert (y_{\Phi})_{e1} \rvert^2$ reaches at the largest value at $\sim 0.1$, while for $\lvert (y_{\Phi})_{\mu 1}\rvert^2$ and $\lvert (y_\Phi)_{\tau 1} \rvert^2$, they could reach values well beyond that for $\lvert (y_{\Phi})_{e1} \rvert^2$. This is a reflection of the loose constraints from tau-relevant processes.\footnote{Although the largest value $(y_\Phi)_{\mu 1}$ and $(y_\Phi)_{\tau 1}$ could reach are much larger than that of $(y_\Phi)_{e 1}$, there still exist parameters with $(y_\Phi)_{e 1}$ larger than $(y_\Phi)_{\mu 1}$ and $(y_\Phi)_{\tau 1}$.}

\section{Dark Matter Detection} \label{sec:DM}

\subsection{Relic Density}

The current experimental result on dark matter relic density is set as $\Omega_{\rm DM} h^2 = 0.1186\pm 0.0020$~\cite{Tanabashi:2018oca}, where $h$ is the Hubble constant in the unit of $100~\rm{km/(s~ Mpc)}$. For an order-of-magnitude estimation, one has
\begin{equation}
\Omega_{\rm DM}h^2\approx \frac{3\times10^{-27} \text{cm}^3 \text{s}^{-1}}
{\langle \sigma v\rangle}.
\end{equation}

\begin{figure}
\centering
	\includegraphics[width=0.9\textwidth]{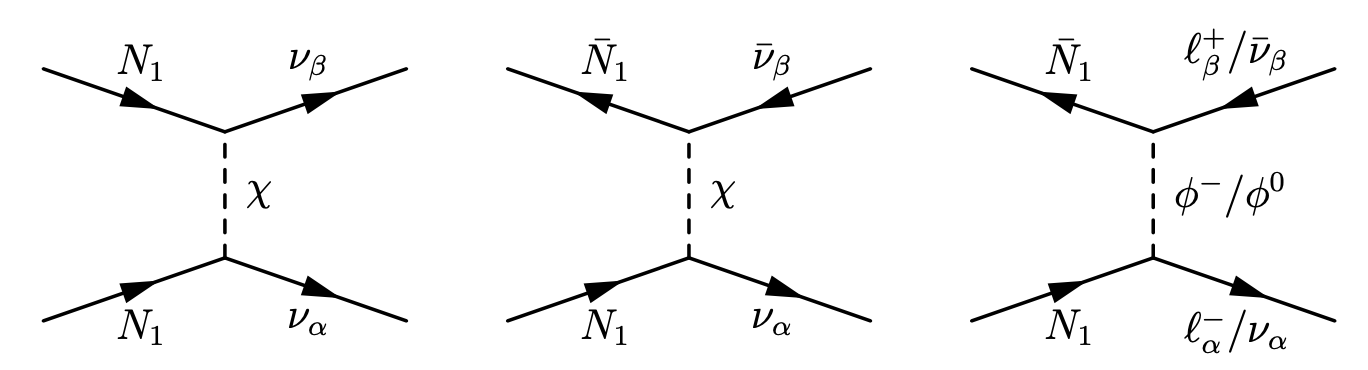}
\caption{Dark matter annihilation in the scotogenic Dirac model.}
\label{fig:dman}
\end{figure}

\begin{figure}
\centering
\includegraphics[width=0.45\textwidth]{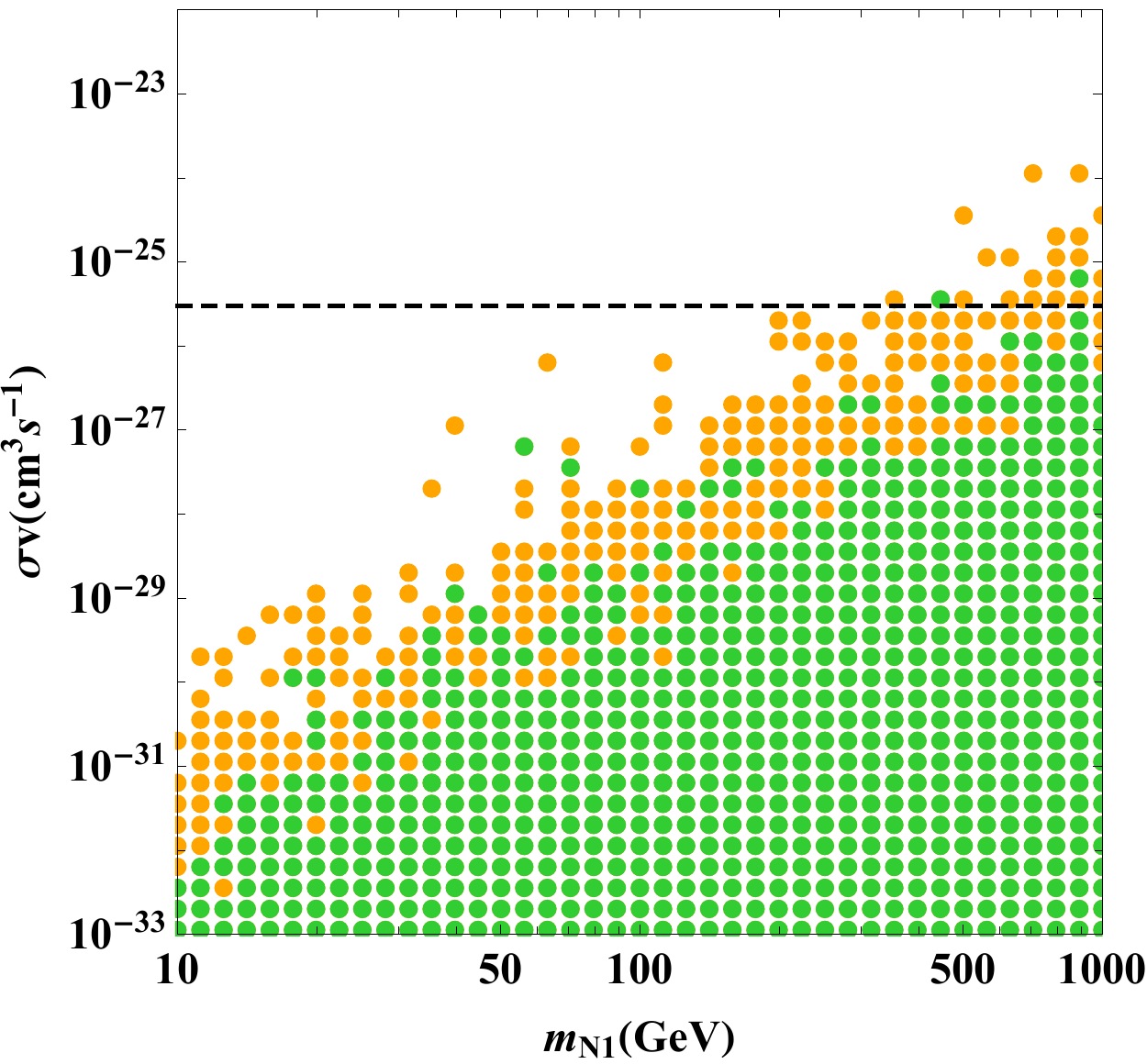}
\caption{DM annihilation rate with contributions from Yukawa coupling $y_\Phi$, the black dashed line represents the dark matter annihilation rate, i.e. $\sigma v = 3\times 10^{-26}~{\rm{cm^{3} s^{-1}}}$, to give out the correct relic density.}
\label{fig:sigvll}
\end{figure}

In this work, we have chosen $N_1$ to act as dark matter candidate, it could annihilate through channels as
\begin{equation}
N_1N_1\to \nu\nu,\ N_1\bar{N_1}\to \nu\bar{\nu},\ N_1\bar{N_1}\to \ell^+\ell^-.
\end{equation}
We have depicted these channels in Figure~\ref{fig:dman}.  The annihilation processes are related to both Yukawa couplings $y_\Phi$ and $y_\chi$. The lepton flavor violating processes have set strict bounds on the matrix elements of $y_\Phi$, which we have discussed in the above section, e.g. see Figure~\ref{fig:yyi1LFV}. For the $\phi$-mediated channels, the cross section reads
\begin{eqnarray}
\sigma(N_1\bar{N}_1\to \ell_\alpha^-\ell_\beta^+) &=& \frac{|(y_\Phi)_{\alpha 1} (y_\Phi^\ast)_{\beta 1}|^2}{32\pi v_{\rm{rel}}} \frac{m_N^2}{(m_N^2+m_{\phi}^2)^2}, \label{eq:annl} \\
\sigma(N_1\bar{N}_1\to \nu_\alpha \overline{\nu}_\beta) &=& \frac{|(y_\Phi)_{\alpha 1} (y_\Phi^\ast)_{\beta 1}|^2}{32\pi v_{\rm{rel}}} \frac{m_N^2}{(m_N^2+m_{\phi}^2)^2},\label{eq:annn}
\end{eqnarray}
from Figure \ref{fig:sigvll} one could see that a big portion of the parameter region is too small to account for the correct annihilation rate to give out the current relic density of dark matter.
This is the same as in the scotogenic Majorana  neutrino model, which has been discussed in~\cite{Vicente:2014wga}. In our work, the dark matter annihilation would mainly through the Yukawa $y_\chi$, and the annihilating final states are neutrinos. The annihilating cross section of processes $N_1N_1\to\nu_\alpha\nu_\beta$ and $N_1\bar{N}_1\to \nu_\alpha\bar{\nu}_\beta$ are given as
\begin{eqnarray}
\sigma(N_1N_1\to\nu_\alpha\nu_\beta) &=&\frac{\lvert(y_\chi)_{\alpha 1} (y_\chi)_{\beta 1}\rvert^2}{16\pi v_{\rm{rel}}} \frac{m_N^2}{(m_N^2+m_\chi^2)^2}, \label{eq:annn1}\\
\sigma(N_1\bar{N}_1\to \nu_\alpha\bar{\nu}_\beta) &=& \frac{|(y_\chi)_{\alpha 1} (y_\chi^\ast)_{\beta 1}|^2}{32\pi v_{\rm{rel}}} \frac{m_N^2}{(m_N^2+m_{\chi}^2)^2},
\label{eq:annn2}
\end{eqnarray}
an additional $1/2$ factor, due to the identical neutrino final states, should be multiplied when $\alpha=\beta$ in process $N_1N_1\to \nu_\alpha\nu_\beta$.

\subsection{Indirect Detection}

Now we switch into the dark matter indirect detection. From the previous discussions, we know the dark matter annihilation final states are either charged leptons or neutrinos, with the annihilation cross section given in Equations \ref{eq:annl}, \ref{eq:annn}, \ref{eq:annn1}, and \ref{eq:annn2}.

\begin{figure}
	\centering
	\includegraphics[width=0.45\textwidth]{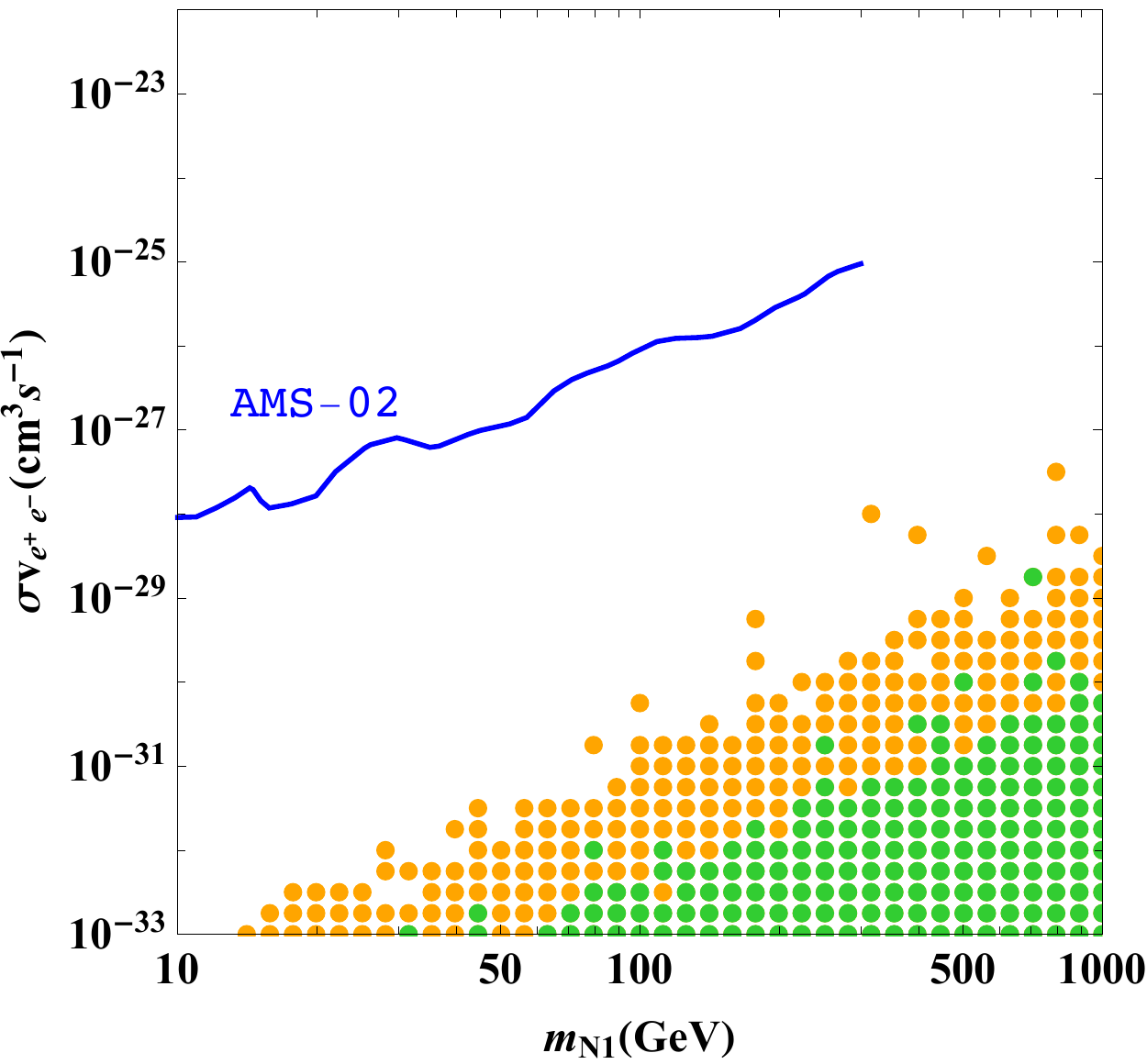}
	\includegraphics[width=0.45\textwidth]{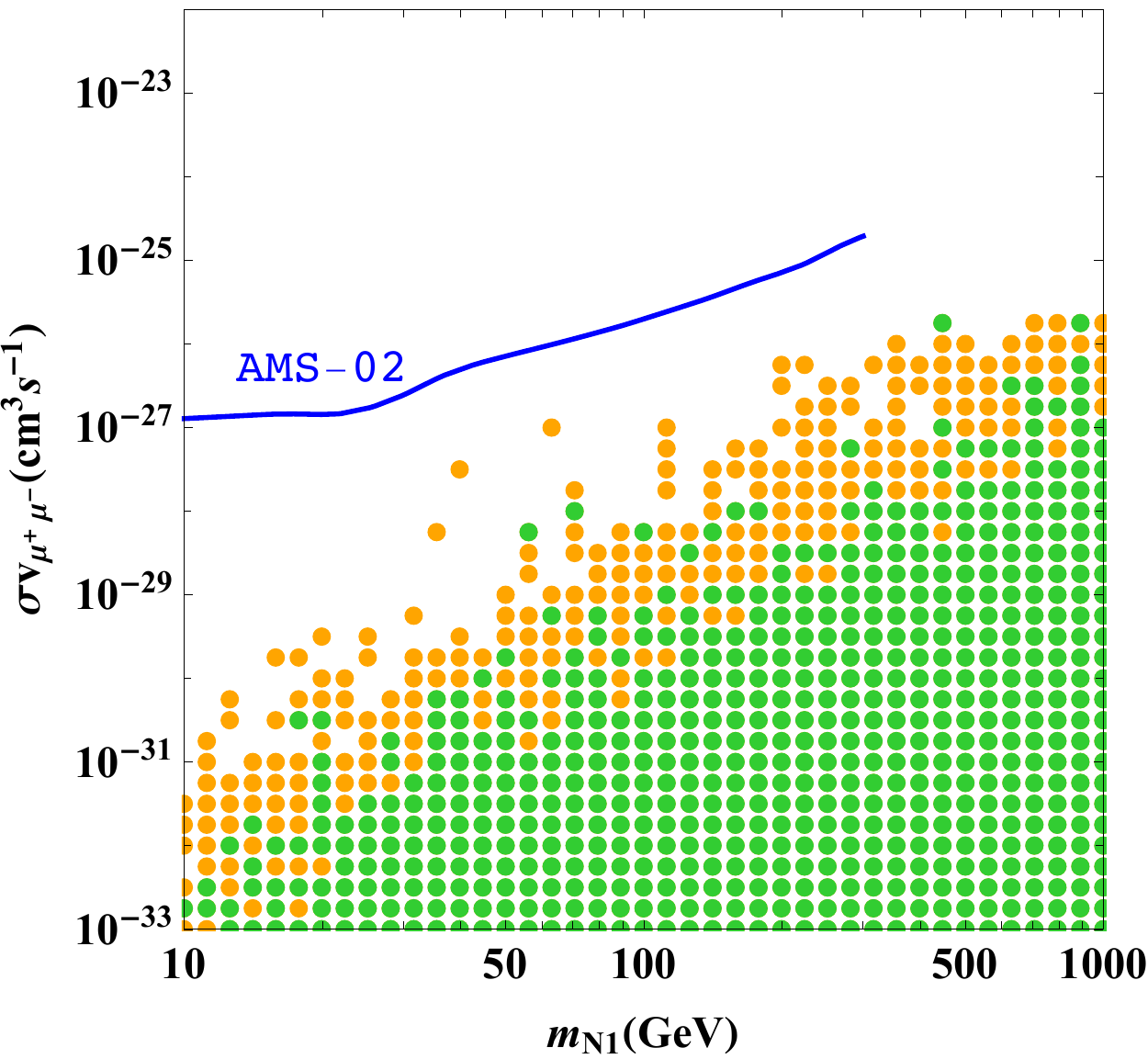}
	\includegraphics[width=0.45\textwidth]{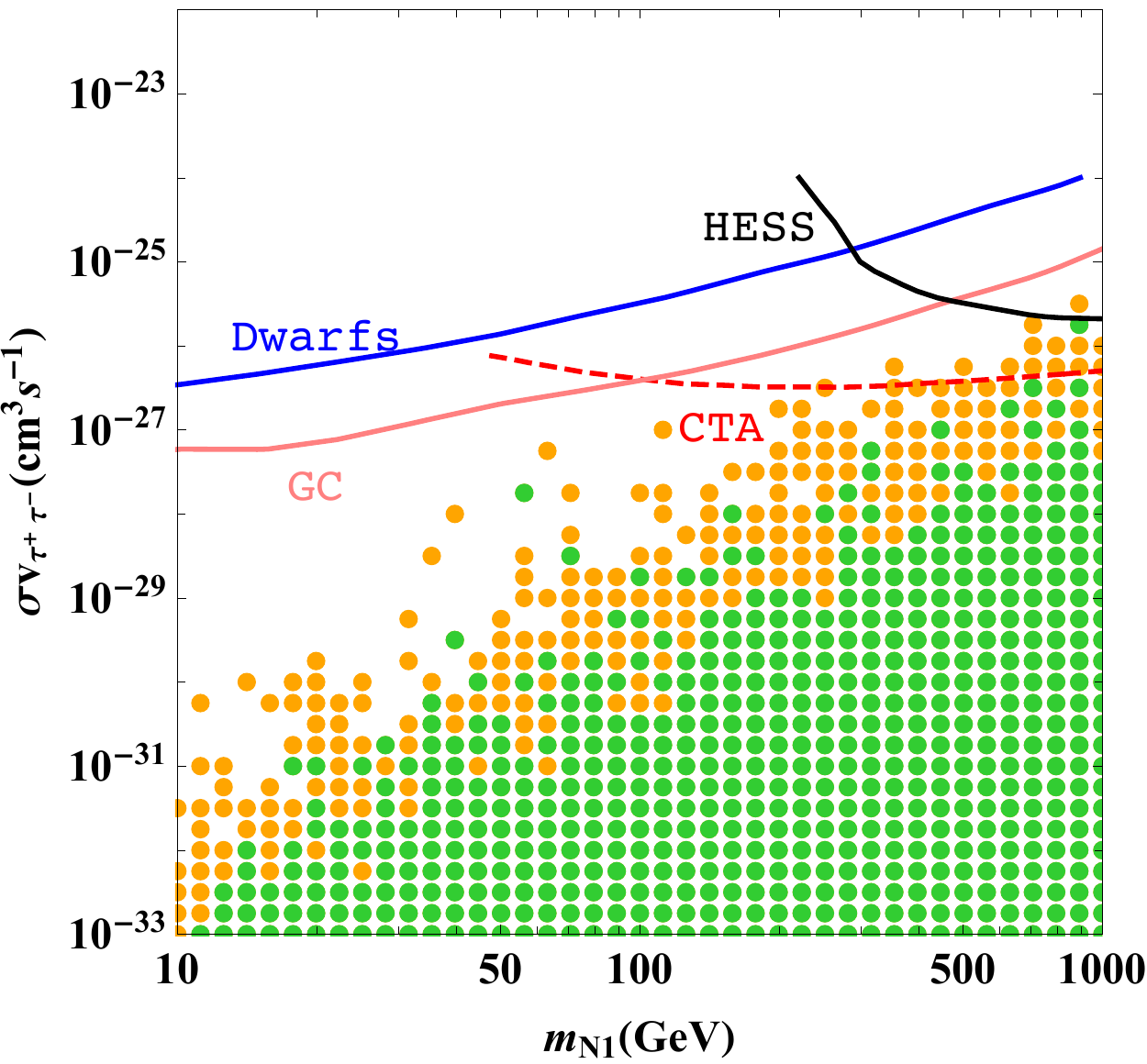}
	\includegraphics[width=0.45\textwidth]{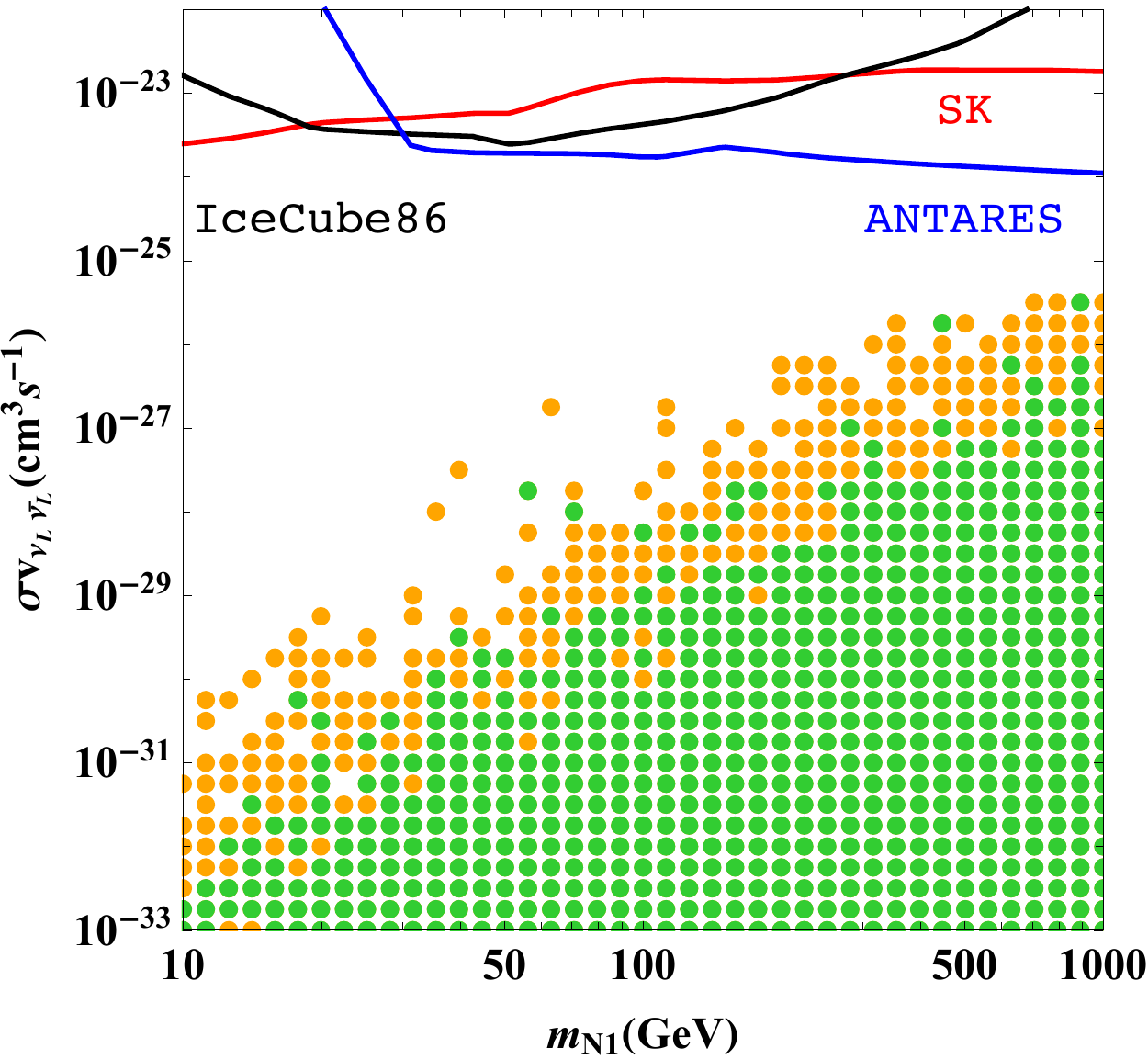}
	\caption{Indirect detection limit from dark matter annihilating into leptons, for annihilating into neutrino final states, only the SM left-handed neutrinos can be detected.}
	\label{fig:ID}
\end{figure}

The recent research~\cite{Abazajian:2020tww} had a discussion on dark matter annihilation to account for the extended excess of gamma ray from the Milky Way Galactic Center, which is observed by Fermi-LAT. Constraints on the annihilation rate of $\tau^+\tau^-$ final states had been set.  We show it in Figure \ref{fig:ID}, where the blue and pink lines stand for upper limits from Fermi-LAT, assuming different dark matter spatial morphologies~\cite{Abazajian:2020tww}. With the dark matter mass increased, the exclusion capabilities will become weaker. Limits from HESS~\cite{Abdallah:2016ygi} and detecting capability of the future experiment CTA~\cite{Carr:2015hta} are also given. They are more sensitive in the large mass region.  The orange(green) dots are those survived from relic density requirement and current(future) LFV constraints. One could see that the relic density and LFV constraints are mostly tighter than indirect detection limits. For the current experiments, Fermi-LAT and HESS assuming various dark matter profiles, their limits are all weaker than the current LFV constraints. For the future experiment, CTA has mild exclusion capability in the large mass region, while it's still weaker than the future LFV experiments. The $e^+e^-$ and $\mu^+\mu^-$ annihilation channel had also been studied in the literature, i.e. Ref.~\cite{Bergstrom:2013jra}, but the exclusion limits are even looser.  Ref.~\cite{Bergstrom:2013jra} had given bounds on the annihilation cross section based on the AMS-02 data, which are shown in Figure \ref{fig:ID}. For the $e^+e^-$ final states the annihilation cross section in this model is pretty small, as the Yukawa $y_\Phi$ has been severely constrained by LFV processes. For the $\mu^+\mu^-$ annihilation channel, constraints on $y_\Phi$ have been eased a little bit and the cross section is comparable to the $\tau^+\tau^-$ channel, but it's still beyond the detecting capability of AMS-02. The dark matter annihilating into neutrinos have also been probed by various experiments, the lower-right subfigure shows the experimental limits on cross section of $\nu_L\bar{\nu}_L$ final states and the corresponding predictions in our model, both Equations \ref{eq:annn} and \ref{eq:annn2} could contribute to the annihilation. Current bounds from Super-K \cite{Abe:2020sbr}, IceCube \cite{Aartsen:2017ulx}, and ANTARES \cite{ANTARES:2019svn} are at most able to exclude $\langle \sigma v\rangle\sim10^{-24}\text{cm}^3\text{s}^{-1}$, which are far beyond  the thermal annihilation cross section, hence cannot exclude any parameters in our model.


\subsection{Direct Detection}

\begin{figure}
	\centering
	\includegraphics[width=0.9\textwidth]{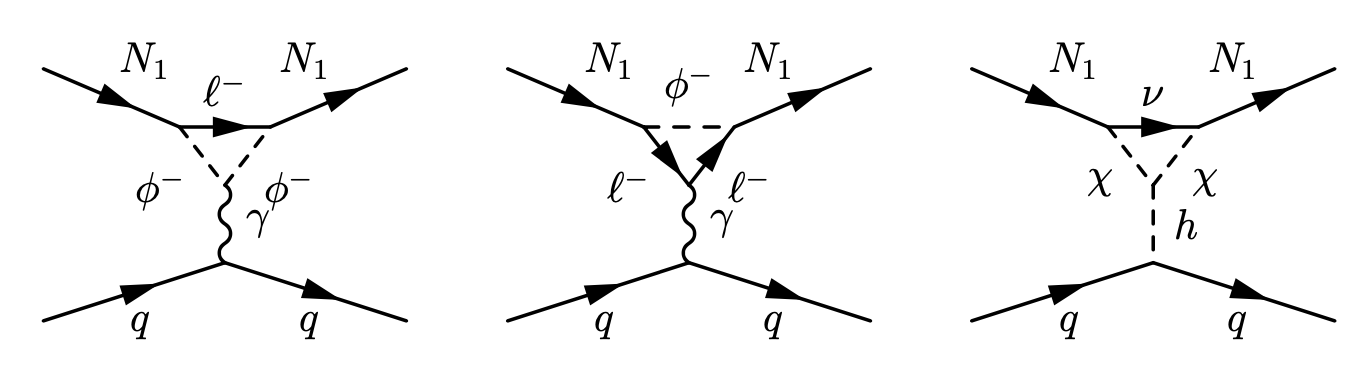}
	\caption{Dark matter scattering on nucleus processes.}
	\label{fig:dmdd}
\end{figure}

Now we discuss the direct detection in this model. As $N_1$ is a neutral $SU(2)_L$ singlet, it has no direct coupling to photon and $Z$ boson. The $Z_3$ symmetry also forbids the direct coupling to Higgs boson. Hence the lowest contribution to direct detection takes place at the one-loop level. From the different dependence on related Yukawa couplings, the direct detection processes could be identified as $y_\Phi$ and $y_\chi$-related processes. The $y_\Phi$-related processes could be further grouped, according to the mediated gauge boson, as $\gamma$, $Z$ and Higgs-mediated. The photon-mediated processes would offer the largest contribution, compared to the $Z$ and Higgs-mediated processes, as the existence of suppression on gauge boson propagator and SM fermion masses. Hence in the $y_\Phi$-related processes, we only concern the photon-mediated processes. The diagrams are depicted in Figure~\ref{fig:dmdd}. For the $y_\chi$-related process, only the Higgs-mediated diagram will contribute. It suffers also the suppression from SM quark masses, but the $y_\chi$ coupling could be large.

We start from the $y_\chi$-related process, which is the last diagram in Figure~\ref{fig:dmdd}. It could be termed into an effective operator form as
\begin{equation}
\mathcal{O}_S = (\bar{N}_1N_1)(\bar{q}q),
\end{equation}
with the Wilson coefficient calculated as
\begin{equation}
C_S= - \sum_\ell\frac{3 \lvert (y_\chi)_{\ell 1} \rvert^2}{16\pi^2 m_h^2 m_{N1}} \frac{\lambda_5 m_q}{4} h\left(\frac{m_{N1}^2}{m_\chi^2}\right).
\end{equation}
The loop function is
\begin{equation}
h(r) = \frac{r+(1-r)\ln(1-r)}{r}.
\end{equation}
The cross section of $N_1$ scattering off from a proton is then calculated as~\cite{Ibarra:2016dlb}
\begin{equation}
\sigma_S = \frac{4}{\pi} \frac{m_{N1}^2 m_p^2}{\left(m_{N1}+m_p\right)^2} m_p^2 \left(\frac{C_S}{m_q}\right)^2 f_p^2,
\end{equation}
with $m_p$ stands for the proton mass and $f_p\approx 0.3$ is the scalar form factor.

\begin{figure}
	\centering
	\includegraphics[width=0.46\textwidth]{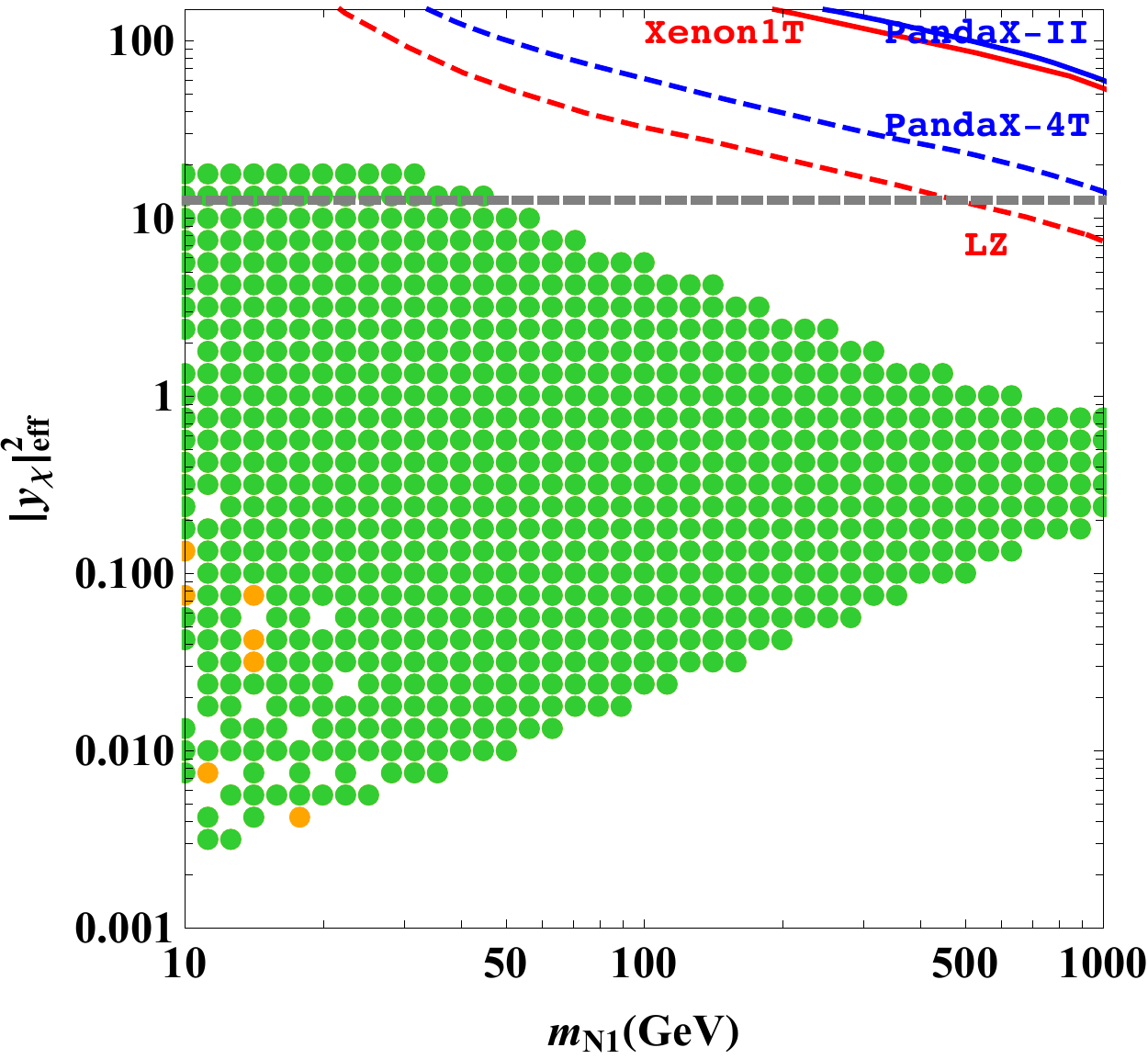}
	\includegraphics[width=0.46\textwidth]{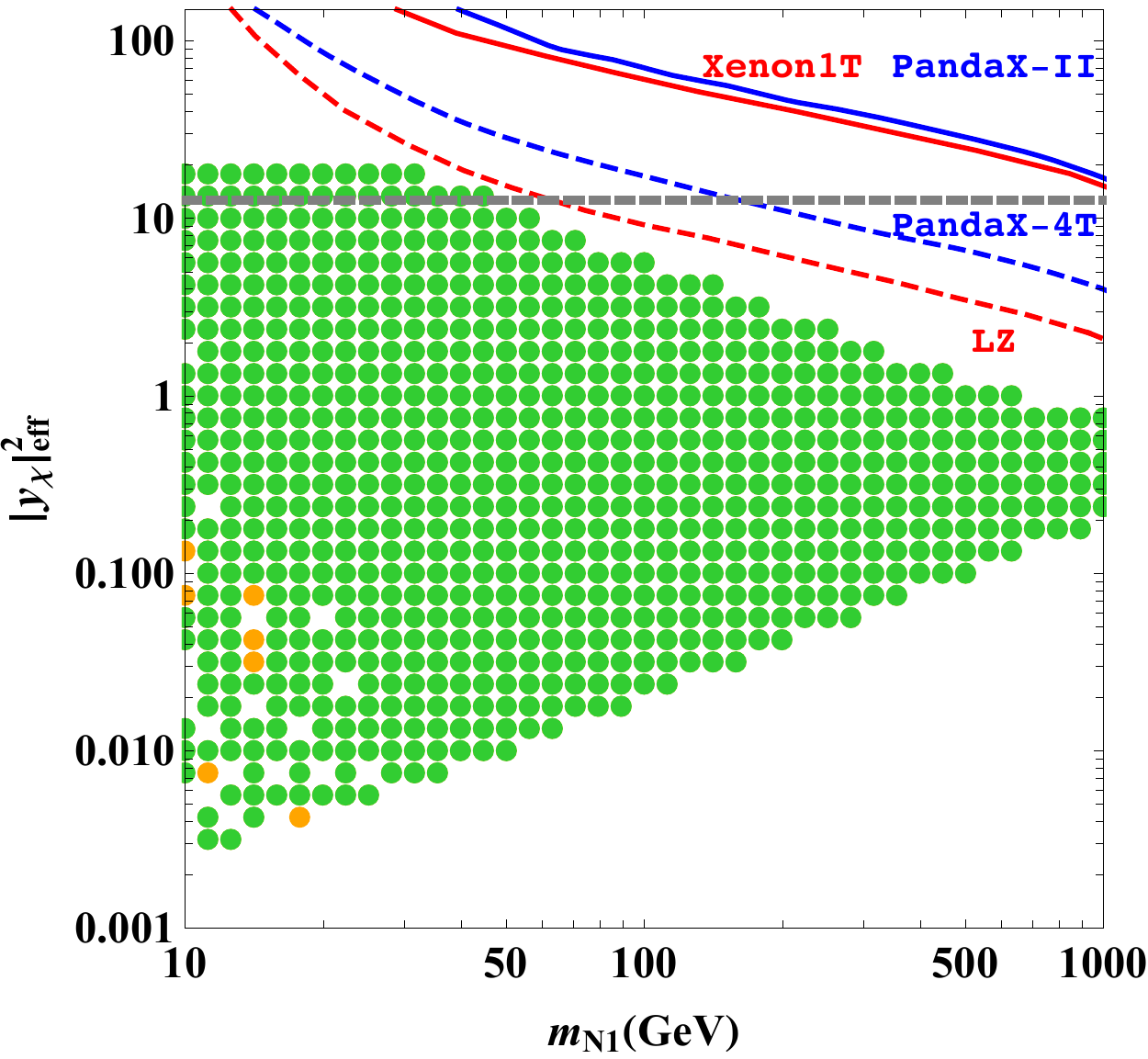}
	\includegraphics[width=0.45\textwidth]{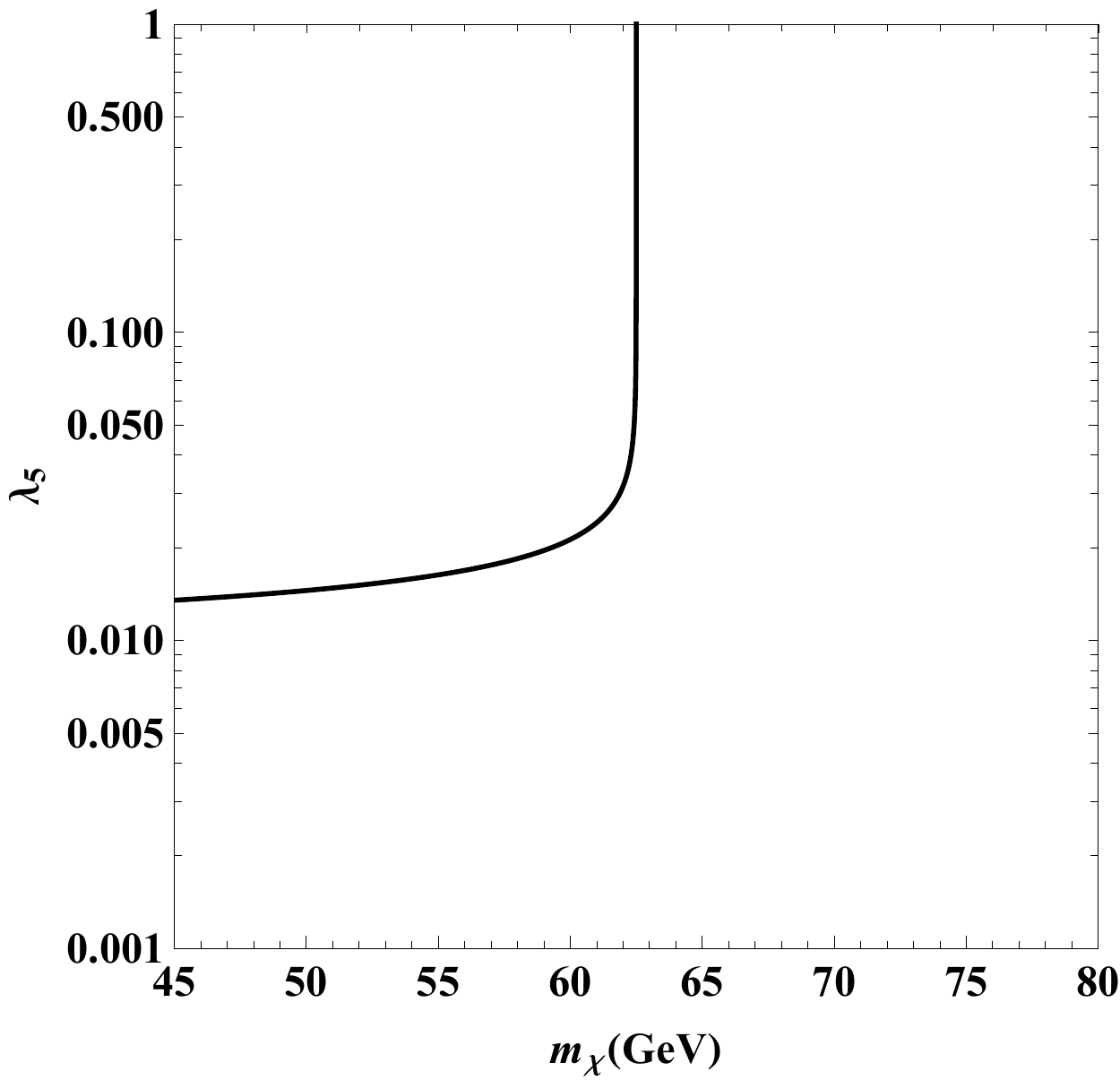}\hspace{0.47\textwidth}
	\caption{Distribution of $|y_\chi|_{\rm eff}^2$ as a function of dark matter mass, the colored lines show bounds from the current(solid) and future(dashed) direct detection experiments, with $m_\chi$ being fixed at $1500~\rm{GeV}$, the gray dashed line stands for the perturbative validity upper limit, i.e. $|y_\chi|_{\rm eff}^2=4\pi$. The difference between the two subfigures is the value of $\lambda_5$, which is fixed as $\lambda_5 = 1$(upper left) and $\lambda_5 = \sqrt{4\pi}$(upper right). Also given is the limit on $\lambda_5$ from Higgs invisible decay, when $m_\chi < m_h/2$. }
	\label{fig:DDch}
\end{figure}

The direct detection results could then be translated into bounds on the $y_\chi$-$m_{N1}$ plane, which we show in Figure~\ref{fig:DDch}. Here we have fixed $m_\chi=1500~\rm GeV$ to give an example, and defined an effective Yukawa $|y_\chi|_{\rm eff}^2= \left(|y_\chi|_{e1}^2+|y_\chi|_{\mu 1}^2+|y_\chi|_{\tau1}^2\right)/3$. For the current direct detection experiments, results are from PandaX-II~\cite{Cui:2017nnn}  with an exposure of $54$~ton-day, and Xenon1T~\cite{Aprile:2018dbl}  with exposure of $1~$ton-year. While for the future detecting capabilities, the experiments PandaX-4T~\cite{Zhang:2018xdp}  with exposure of $5.6~$ton-year, and LZ~\cite{Akerib:2018dfk}  with exposure of $5.6\times 1000~$ton-day, are used. The two sub-figures are different in the value of $\lambda_5$, with $\lambda_5 = 1$ on the left-hand side and $\lambda_5 = \sqrt{4\pi}$ on the right-hand side. We see that bounds from the direct detection on $y_\chi$ are rather loose. A common relation between $y_\chi^2$ and dark matter mass $m_{N1}$ is that, with larger dark matter mass the upper limit on $y_\chi^2$ will be stricter. That is because the direct detection process is Higgs-mediated, which will flip the chirality of dark matter, making the cross section proportional to the dark matter mass. The $y_\chi$ is bounded by the relic density observed today, as the dark matter annihilation is dominant through the $\chi$-mediated processes. Meanwhile, the two Yukawas $y_\chi$ and $y_\Phi$ are connected to each other through the neutrino mass in Equation \ref{eq:numass}. The $y_\chi$ is then indirectly constrained by LFV. We show the colored points that satisfy the requirement of giving the correct dark matter relic density while at the same time surviving from current and future LFV constraints. The orange dots lie beyond the current LFV experiments detecting capabilities while could be detected by the future experiments, the green dots are beyond even the future experiments detecting capabilities. The gray dashed lines indicate the perturbative validity upper limit $4\pi$. One could find that the LFV, relic density, and perturbativity requirements together give out constraints tighter than that from direct detections. Moreover, if the scalar $\chi$ is light enough, it will then induce the Higgs invisible decay. The experiment results would set constraints on the $\lambda_5-m_\chi$ plane. We show the result in Figure \ref{fig:DDch} and will make a precise discussion later. One could see that for a light $m_\chi$, i.e. $m_\chi<m_h/2$, the upper limit on $\lambda_5$ has been set to approach $\sim 0.01$, for such small $\lambda_5$ the direct detection constraints would become even looser.

Now we switch into the $y_\Phi$-related processes. The photon diagrams, after integrating out the heavy degree of freedoms in the loop, will contribute to the vector current and magnetic dipole operators:
\begin{eqnarray}
\mathcal{O}_{\rm{VV}} &=& \left(\bar{N}_1\gamma^\mu N_1\right) \left(\bar{q} \gamma_\mu q\right) \\
\mathcal{O}_{\rm{mag.}} &=& \frac{e}{8\pi^2}\bar{N}_1\sigma^{\mu\nu} N_1 F_{\mu\nu}.
\end{eqnarray}
It is the most obvious difference from the scotogenic Majorana model, since these two operators are vanished for $N_1$ to be a Majorana neutrino. Hence it's natural to expect a larger cross section compared to the Majorana neutrino model. The Wilson coefficients for these two operators, up to the lowest order of momentum transfer, are calculated as
\begin{eqnarray}
C_{\rm VV} &=& \frac{e^2 Q_q}{16\pi^2} \sum_{\ell} \frac{|(y_\Phi)_{\ell 1}|^2}{12 m_\phi^2 x_{N1}^4} \frac{1}{x_{N1}^4 + (1-x_\ell^2)^2 - 2x_{N1}^2(1+ x_\ell^2)} \\
&\times&
\begin{pmatrix}
\scriptstyle
-2 x_{N1}^2( 3x_{N1}^4 - x_{N1}^2(7+5 x_\ell^2) + 4 (1- x_\ell^2)^2) \\
\scriptstyle
- \left(x_{N1}^6 + x_{N1}^4 (6-10 x_\ell^2) + x_{N1}^2 (-15 -2 x_\ell^2 + 17 x_\ell^4) + 8(1- x_\ell^2)^3 \right) \ln x_\ell \notag\\
\scriptstyle
-\left(3 x_{N1}^8 - x_{N1}^6 (13 + 11 x_\ell^2) + x_{N1}^4 (25 - 2 x_\ell^2 + 25 x_\ell^4)\right. \\
\scriptstyle
\left.- x_{N1}^2 (1 - x_\ell^2)^2 (23 + 25 x_\ell^2) + 8 (1 - x_\ell^2)^4 \right) g(x_{N1}, x_\ell)
\end{pmatrix}\\
\mu_{\rm mag.} &=& -\sum_{\ell}\frac{|(y_\Phi)_{\ell 1}|^2}{8 m_\phi x_{N1}^3} \left[ x_{N1}^2 + (1- x_\ell^2) \ln x_\ell \right.\\ \nonumber
&& \left.+ \left((1-x_\ell^2)^2 - x_{N1}^2 (1+x_\ell^2)\right) g(x_{N1}, x_\ell)\right]
\end{eqnarray}
with $x_{N1} = m_{N1}/m_\phi, x_\ell = m_\ell/m_\phi$, and the loop function
\begin{equation}
g(x_{N1}, x_\ell) = \frac{\ln \left(\frac{1- x_{N1}^2 + x_\ell^2 + \sqrt{x_\ell^4 + (1-x_{N1}^2)^2 - 2 (1+ x_{N1}^2) x_\ell^2}}{2x_\ell}\right)}{\sqrt{x_\ell^4 + (1-x_{N1}^2)^2 - 2 (1+ x_{N1}^2) x_\ell^2}}.
\end{equation}
Our calculations are consistent with the results in Ref.~\cite{Herrero-Garcia:2018koq}. For the case of electron running in the loop, the approximation of $m_\ell \gg \sqrt{-q^2}$ is not appropriate, Ref.~\cite{Herrero-Garcia:2018koq} has given an alternative expression.
 \begin{figure}
\centering
\includegraphics[width=0.45\textwidth]{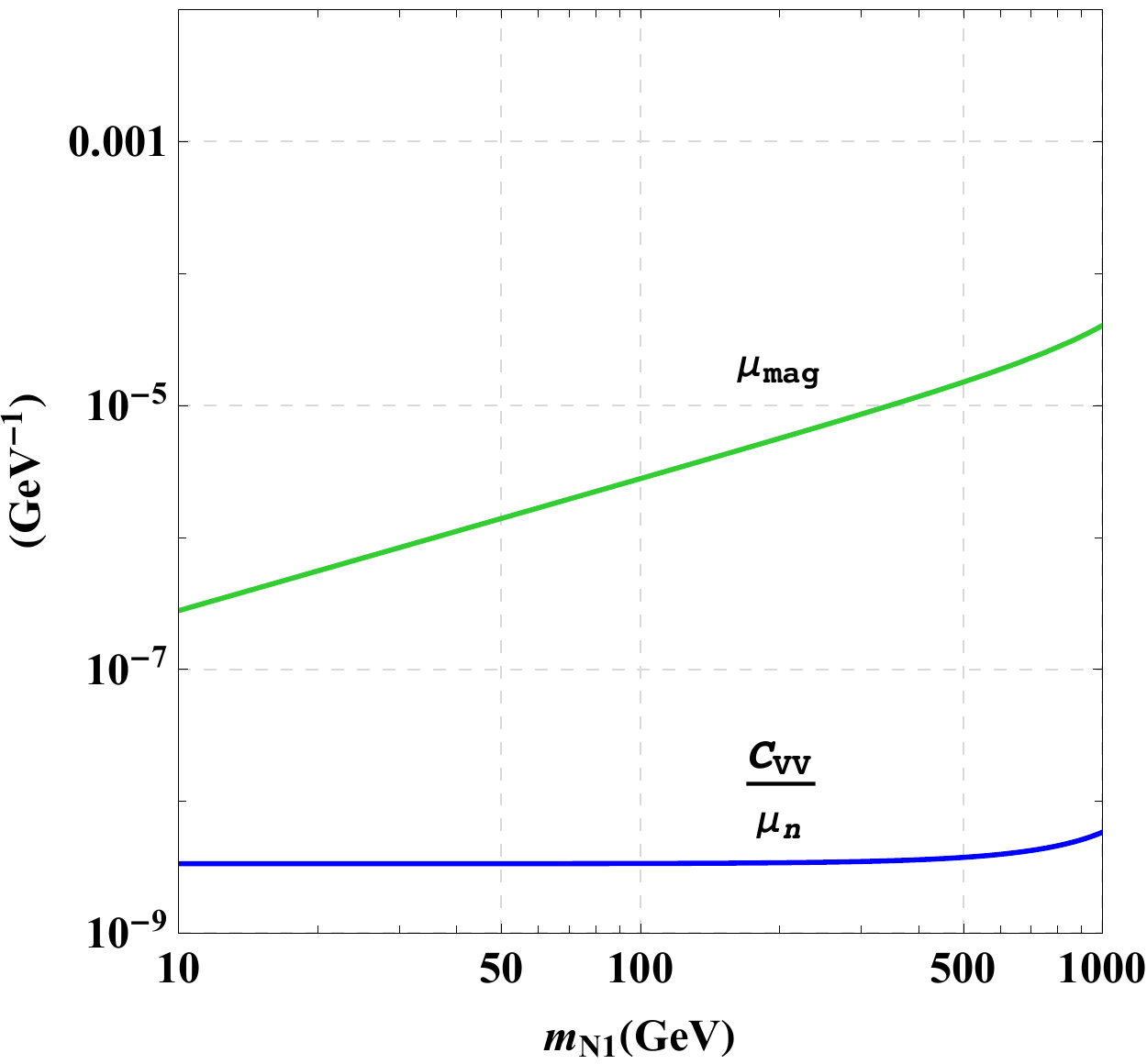}
\caption{Wilson coefficients of the two operators contributing to $y_\Phi$-related direct detection processes, with $m_\phi=1500~\rm{GeV}$, the vector current coefficient has been normalized by nuclear magneton.}
\label{fig:dpvv}
\end{figure}
In Figure \ref{fig:dpvv} we present the comparison between these two Wilson coefficients, with $C_{\rm VV}$ being normalized by nuclear magneton $\mu_N= e/2m_p$($m_p$ stands for the mass of proton). We have fixed $m_\phi=1500~\rm{GeV}$ here. One could see that the dipole coefficient is always much larger than the vector coefficient, and the former one increased as the mass of dark matter being larger. This is because the dipole operator is chirality-flipped, which means the coefficients must be proportional to the mass of dark matter.

To calculate the DM-nucleus scattering event rate, one could match the relativistic DM-quark operators into non-relativistic DM-nucleus operators~\cite{Bishara:2017nnn}. Then the event rate is calculated as~\cite{Anand:2013yka}
\begin{equation}
\frac{d R}{d E_R} = \frac{\rho_{N1}}{m_{N1} m_A} \int_{v_{\rm min}} \frac{d \sigma}{d E_R} v f_{\rm det} (\vec{v}) d^3 v.
\end{equation}
Here $\rho_{N1}$ is the local density of dark matter, $m_A$ is the mass of the nucleus, $f_{\rm det}(\vec v)$ is the dark matter velocity distribution in the detector rest frame, $v_{\rm min}$ is the minimum dark matter velocity to produce a recoil $E_R$, $v_{\rm min} = \sqrt{E_R m_A/\mu_{A N1}}$, with $\mu_{A N1}$ denoting the reduced mass of the DM-nucleus system. The differential cross section is expressed in the form of~\cite{Anand:2013yka,Bishara:2017pfq}
\begin{eqnarray}
	\frac{d \sigma}{d E_R} = \frac{m_A}{2\pi v^2} \frac{4\pi}{2 J_A + 1} \sum_{\tau,\tau^\prime = \{0,1\}} & &\Big[R_M^{\tau\tau^\prime} W_M^{\tau\tau^\prime} (|\vec q|) + R_{\Sigma^{\prime\prime}}^{\tau\tau^\prime} W_{\Sigma^{\prime\prime}}^{\tau\tau^\prime} (|\vec q|) + R_{\Sigma^{\prime}}^{\tau\tau^\prime} W_{\Sigma^{\prime}}^{\tau\tau^\prime} (|\vec q|) \notag \\
	& &+ \frac{|\vec q|^2}{m_{N1}^2}\left(R_\Delta^{\tau\tau^\prime} W_\Delta^{\tau\tau^\prime} (|\vec q|) + R_{\Delta\Sigma^\prime}^{\tau\tau^\prime} W_{\Delta\Sigma^\prime}^{\tau\tau^\prime} (|\vec q|) \right)\Big],
\end{eqnarray}
with the non-relativistic operators encoded in $R$s, the $W$s stand for the nuclear response functions. In the numerical estimation of the direct detection event rate, we use $\mathtt{DirectDM}$~\cite{Bishara:2017nnn} to match the DM-quark operators $\mathcal{O}_{VV}$ and $\mathcal{O}_{\rm mag.}$ into non-relativistic DM-nucleus operators, then $\mathtt{DMFormFactor}$~\cite{Anand:2013yka} is used to numerically calculate the event rate using the non-relativistic operators.

\begin{figure}
	\centering
	\includegraphics[width=0.45\textwidth]{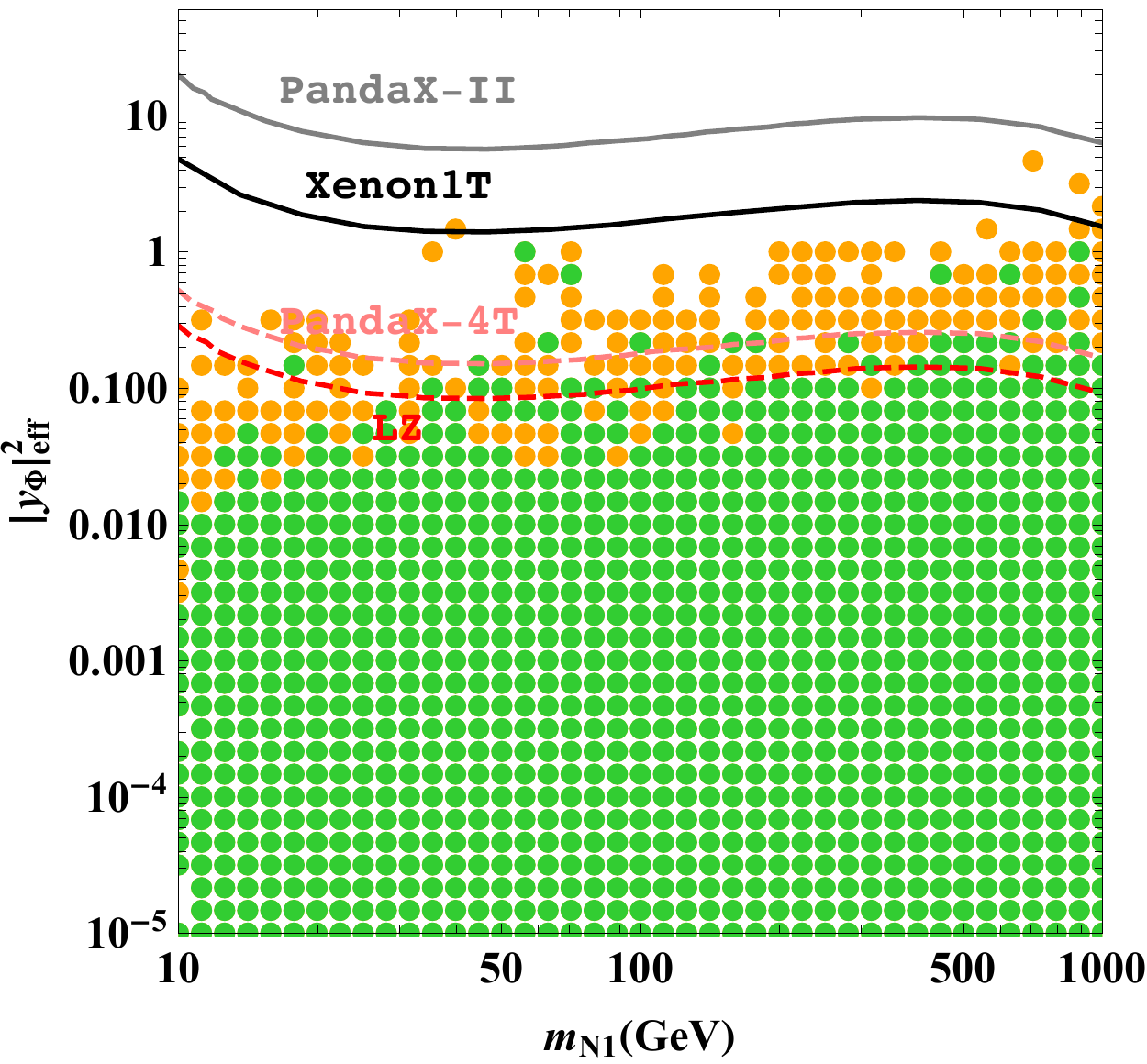}
	\caption{Exclusion limit on the plane of $y_\Phi$-$m_N$ from the current direct detection experiments, where we have set $m_\phi=1500~\rm{GeV}$, the effective $y_\Phi$ is defined as $|y_\Phi|_{\rm eff}^2=(y_\Phi^\dagger y_\Phi)_{11}/3 = \left(|(y_\Phi)_{e1}|^2+|(y_\Phi)_{\mu 1}|^2+|(y_\Phi)_{\tau1}|^2\right)/3$.}
	\label{fig:DD}
\end{figure}

We show our results in Figure~\ref{fig:DD}. The direct detection bounds from experiments have been translated into an upper limit on $90\%$ confidence level of the event number. For these limits, we have fixed $m_\phi=1500~\rm{GeV}$ to show as an example. The colored dots are those survived from dark matter relic density requirement and the LFV constraints, orange dots have passed constraints from current lepton flavor violating constraints and under the future experiments detecting capabilities, while the green dots are beyond the detecting capabilities of the future experiments. Here we have defined an effective Yukawa $|y_\Phi|_{\rm eff}^2= \left(|y_\Phi|_{e1}^2+|y_\Phi|_{\mu 1}^2+|y_\Phi|_{\tau1}^2\right)/3$. We see that most of the parameters which survived from LFV and relic density constraints are located under the bounds from current experiments, though there are still some points that can reach the Xenon1T detecting limit. The future experiments detecting capabilities are also given, which will be improved due to larger amounts of exposures. For larger dark matter mass the nuclear recoil would be increased, while the experiment detecting efficiency will, however, be decreased. But one could see that on the large $m_{N1}$ region, the limit on $|y_\Phi|^2_{\rm{eff}}$ has not been weakened too much, which could be understood from the following discussion: the direct detection is dominated by the dipole operator $\mathcal{O}_{\rm mag.}$, which is related to chirality-flipping of dark matter, the Wilson coefficient $\mu_{\rm mag}$ would then be proportional to dark matter mass. We have shown it in Figure~\ref{fig:dpvv}, hence the cross section will increase for heavier dark matter. This effect makes the excluding limit on $|y_\Phi|^2_{\rm{eff}}$ not be weakened too much by the decreasing detecting efficiency.

\begin{figure}
	\centering
	\includegraphics[width=0.47\textwidth]{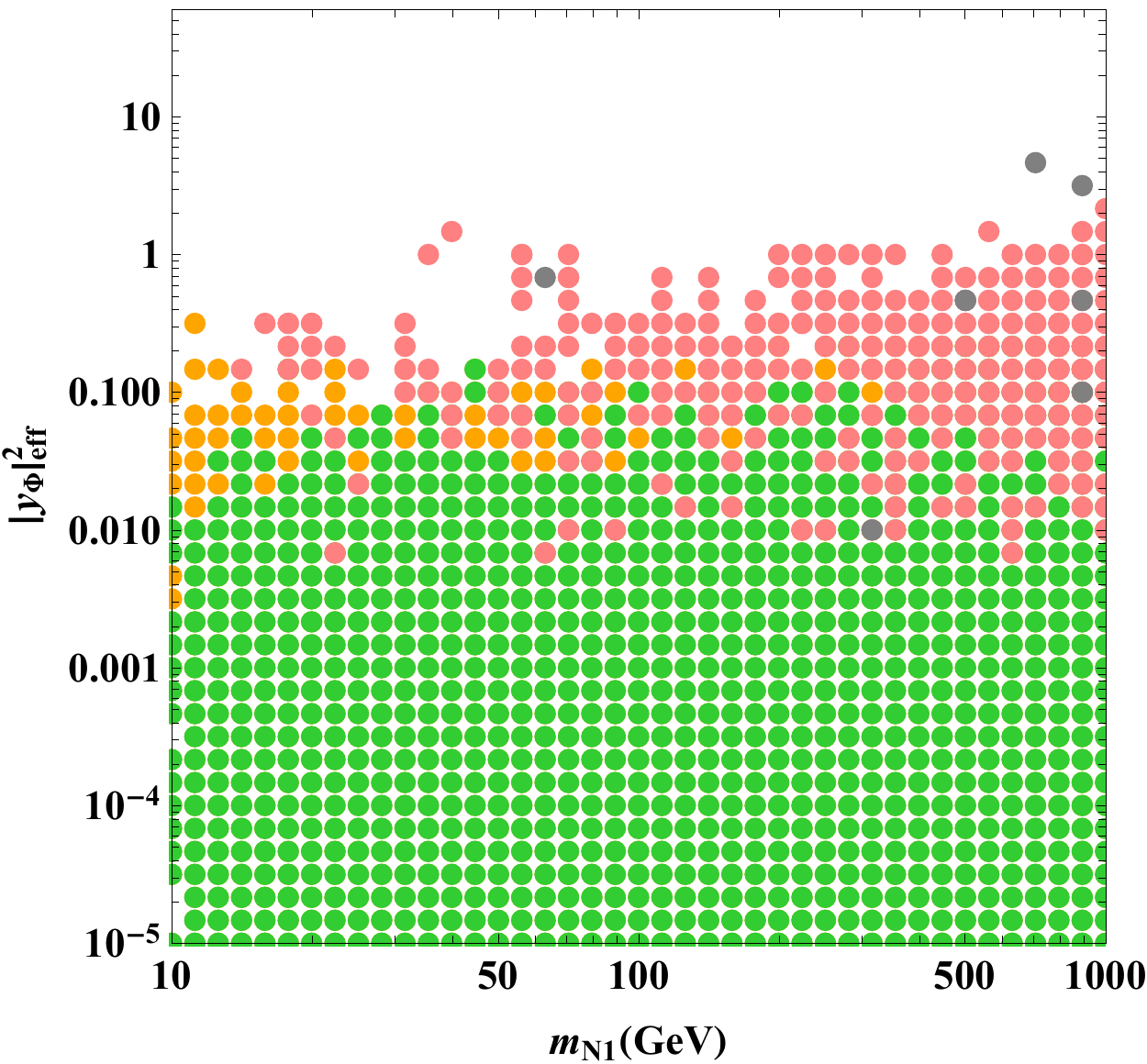}
	\includegraphics[width=0.45\textwidth]{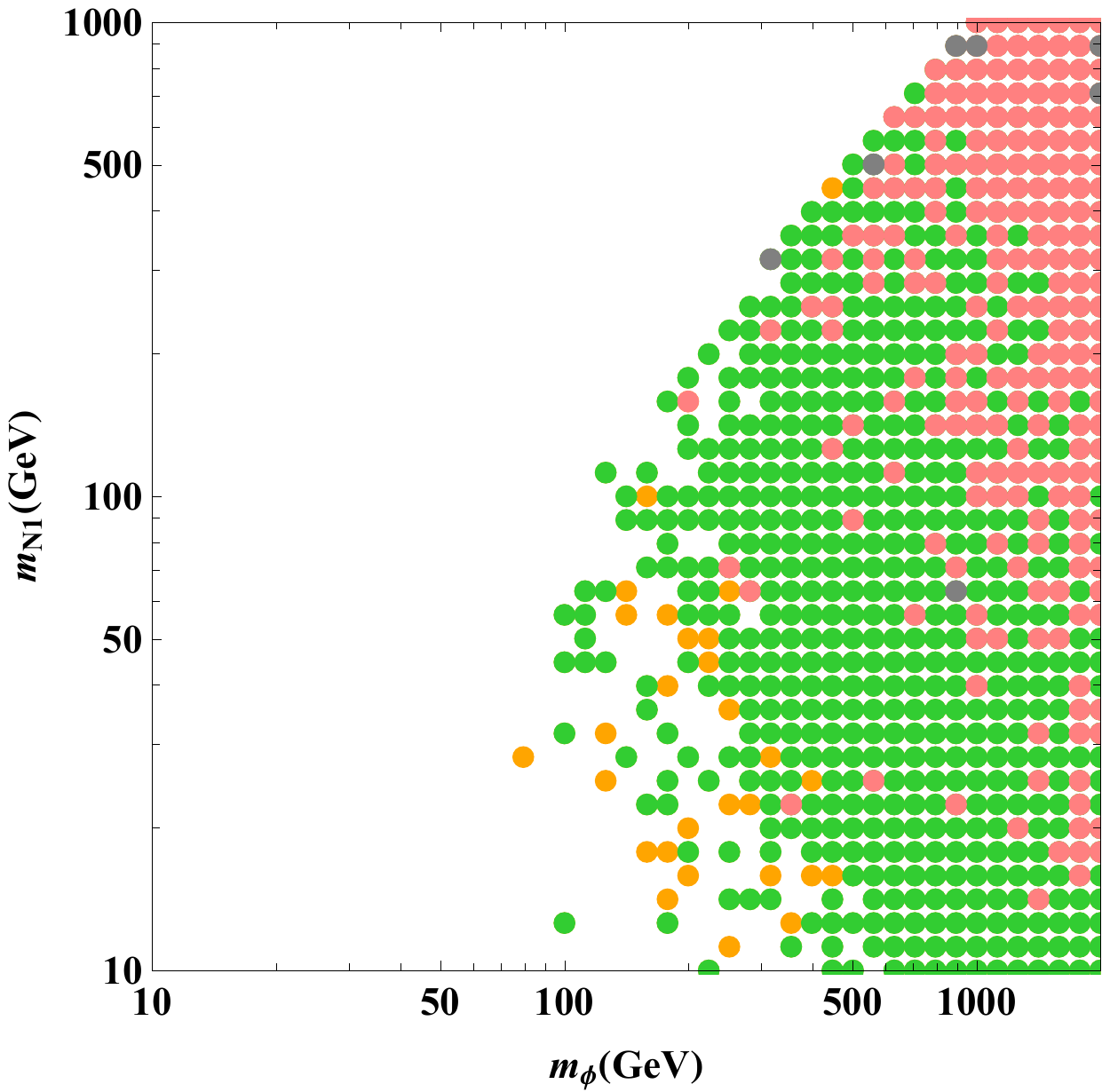}
	\caption{Excluding capabilities of current and future direct detection experiments, after asking for correct dark matter relic density and satisfying the LFV constraints.}
	\label{fig:DDcrtnft}
\end{figure}
In Figure~\ref{fig:DDcrtnft}, we have numerically calculated the direct detection event rates for all the parameters survived from LFV and relic density constraints. Direct detection bounds have been shown in each sub-figures. The gray dot is excluded by current experiments, i.e. PandaX-II and Xenon1T, dots in pink are in the future detecting capabilities of PandaX-4T and LZ, the orange and green dots both satisfy the relic density requirement and are different in that they survive from the current and future LFV constraints, respectively. We have presented our results on $|y_\Phi|_{\rm eff}^2-m_{N1}$ and $m_{N1}-m_\phi$ planes, from the distribution on the former plane we could confirm the comments concluded in the specific $m_\phi=1500~\rm{GeV}$ benchmark point: the current PandaX-II and Xenon1T experiments could barely provide further excluding capabilities beyond the current LFV and relic density constraints. It may be improved by the future experiments PandaX-4T and LZ, with larger amounts of exposures. The large mass region has also been bounded, and even more, severely as the dipole interaction will contribute an enhancement with heavier dark matter. This is also shown in the distribution on $m_{N1}-m_\phi$ plane, where one could find that the direct detections excluded parameters are more concentrated in the large mass region.

\section{Collider Signature}\label{sec:LHC}

\begin{figure}
	\centering
	\includegraphics[width=0.46\textwidth]{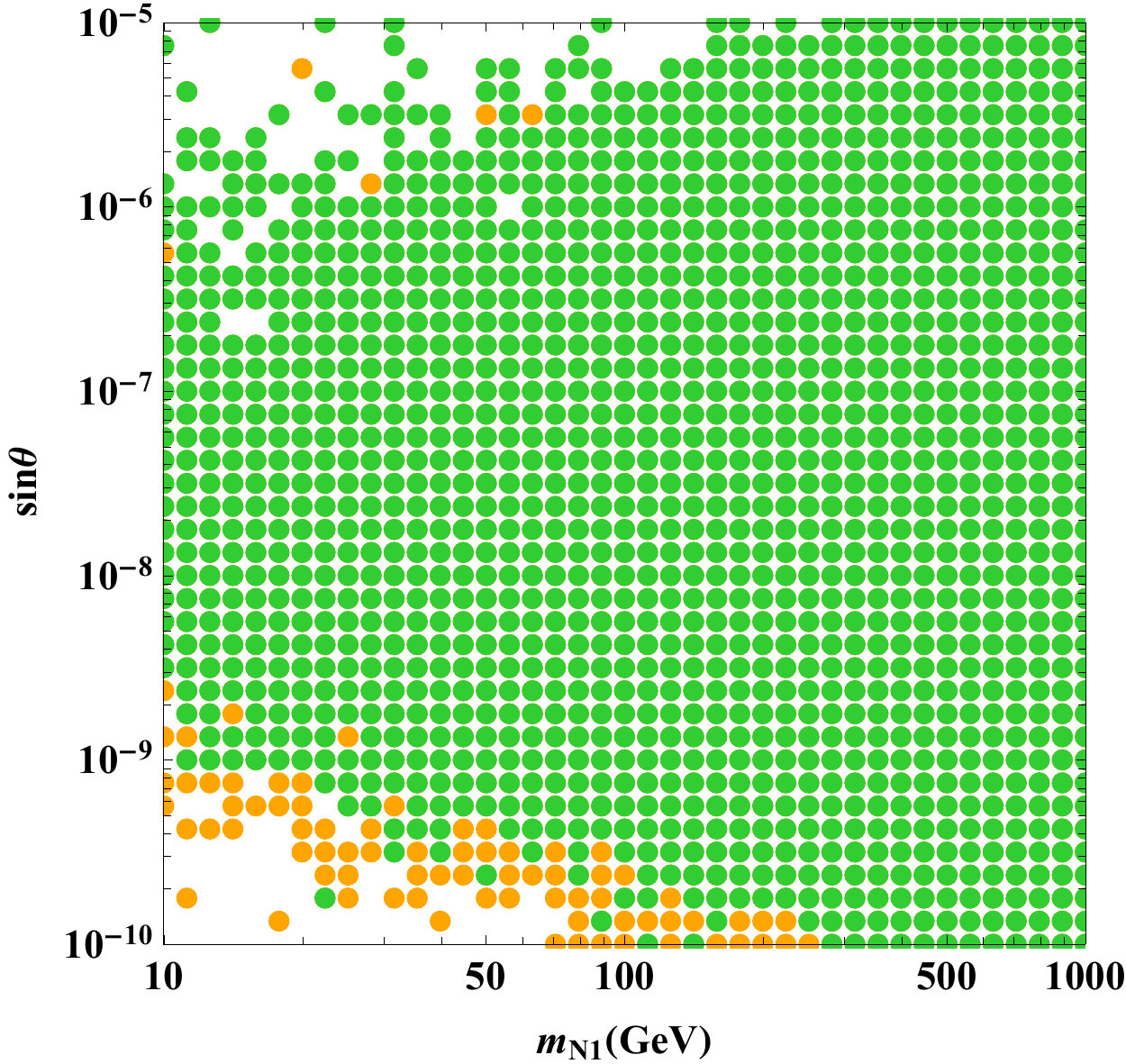}
	\includegraphics[width=0.46\textwidth]{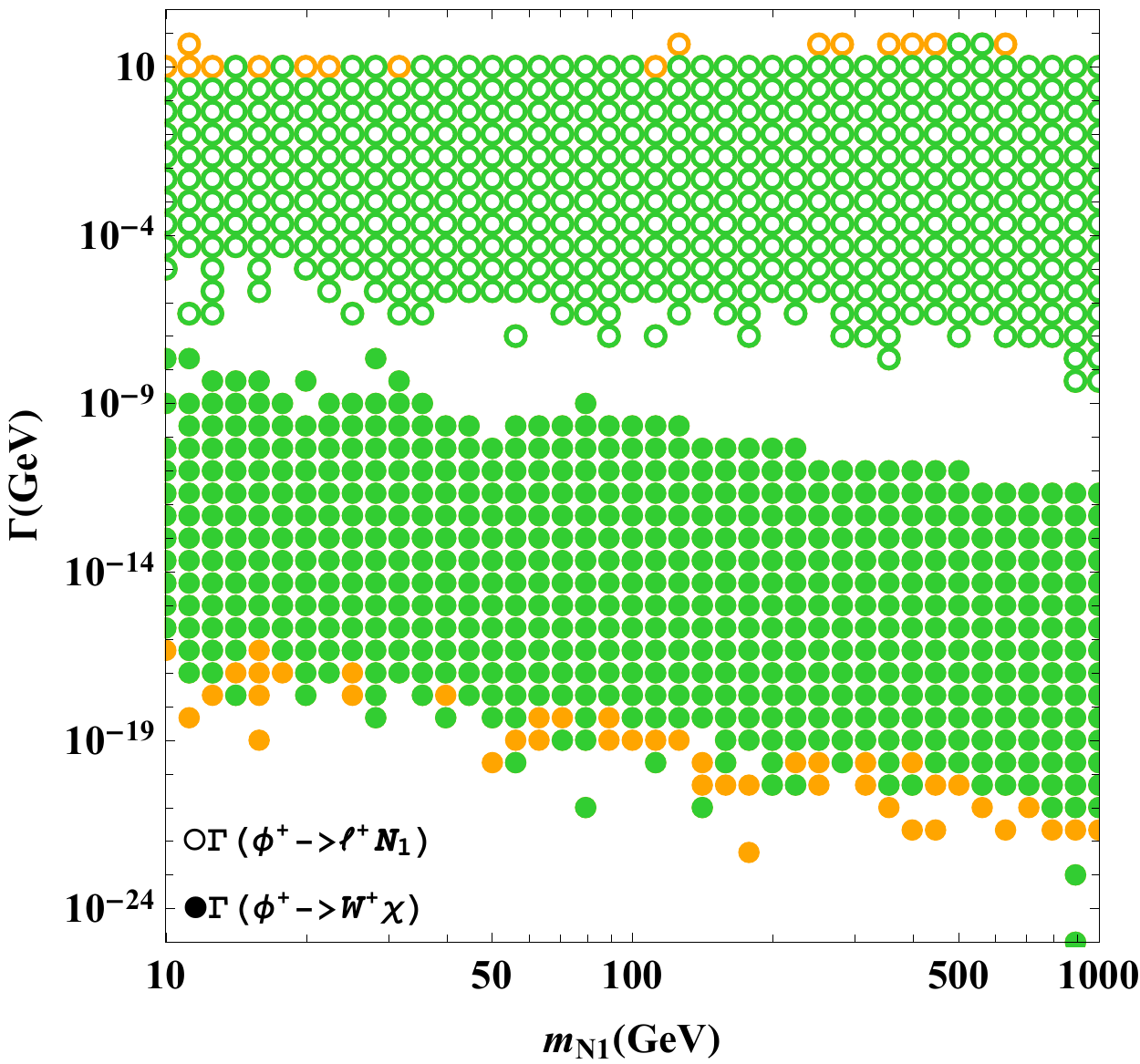}
	\caption{Distribution of the mixing angle $\sin\theta$ and decay widths of the two decay channels in Equations \ref{eq:decay1} and \ref{eq:decay2}, after satisfying LFV, dark matter relic density, direct and indirect detections constraints.}
	\label{fig:br}
\end{figure}

In this section, we will make a brief discussion of the collider signatures in this model. The singlet scalar $\chi$ would decay into neutrinos and the dark matter $N_1$, final states are invisible on colliders. When $\chi$ is lighter than $m_h/2$, the decay channel $h\to \chi \chi\to N_1\bar{\nu} N_1\bar{\nu}$ will contribute to invisible decay, with decay width of
\begin{equation}
	\Gamma_{\rm inv} = \frac{\lambda_5^2 v^2}{32\pi m_h} \sqrt{1-\frac{4m_\chi^2}{m_h^2}}.
\end{equation}
The branching ratio of Higgs invisible decay is then ${\rm BR}_{\rm inv} = \Gamma_{\rm inv}/(\Gamma_{\rm inv}+ \Gamma_{\rm SM})$, with the SM width $\Gamma_{\rm SM} = 4.07~\rm{MeV}$. The latest search with vector boson fusion produced Higgs, using $139~{\rm fb^{-1}}$ collision data with $\sqrt{s}=13~{\rm{TeV}}$, had set the upper limit of invisible decay branching ratio to $0.13$  at $95\%$ confidence level, by ATLAS collaboration~\cite{ATLAS:2020cjb}. While CMS set the upper limit to $0.33$ using data with integrated luminosity of $35.9~{\rm fb^{-1}}$. The limit lowered to $0.19$ with the combination of data from $\sqrt{s}=7$, $8$, and $13~{\rm TeV}$ by Higgs production via gluon fusion, in association with vector boson and vector boson fusion~\cite{Sirunyan:2018owy}. The combination analysis by ATLAS set the upper limits to $0.26$, with data from $\sqrt{s}=7$, $8~\rm{TeV}$, and the $\sqrt{s}=13~{\rm TeV}$ data with integrated luminosity of $36.1~\rm{fb^{-1}}$ by the Higgs production via vector boson fusion and in association with vector boson~\cite{Aaboud:2019rtt}. We have shown the strongest limit on the $\lambda_5-m_\chi$ plane in Figure \ref{fig:DDch}. When $\chi$ is heavy enough, the decay of $\chi \to h \phi_R$ may open, but the decay rate would be suppressed due to the smallness of parameter $\mu$.

The charged component of $\Phi$ could decay into SM charged leptons, i.e. $\phi^+\to \ell^+ N_1$, another decay channel is $\phi^+\to W^+ \chi$, which is originated from the mixing between $\Phi$ and $\chi$, but it is suppressed by the mixing angle, which would be clearer if we write out the two decay widths:
\begin{eqnarray}
\Gamma_{\phi^+\to \ell^+ N_1} &=& \frac{\lvert (y_\phi)_{\ell 1}\rvert^2}{16\pi} \frac{\left(m_\phi^2 - m_{N1}^2\right)^2}{m_\phi^3}, \label{eq:decay1}\\
\Gamma_{\phi^+\to W^+ \chi} &=& \frac{\alpha_{\rm{EM}} \sin^2\theta}{16 \sin^2\theta_w}  \frac{\left[m_\chi^4 + (m_\phi^2 - m_W^2)^2 - 2m_\chi^2(m_\phi^2 + m_W^2)\right]^{3/2}}{m_W^2 m_\phi^3},
\label{eq:decay2}
\end{eqnarray}
where we have neglected the lepton masses, and $\theta_w$ labels the Weinberg angle. To be precise, we present the distribution of $\sin\theta$ and decay widths of the two decay channels in Figure \ref{fig:br}. One could see that the gauge boson decay channel would always be subdominant, due to the large suppression from mixing angle $\theta$. For the case of $m_\phi>m_{N2,3}$, the scalar will also decay into $\ell^+ N_{2,3}$, and $N_{2,3}$ subsequently decay into leptons and $N_1$.

\begin{figure}
	\centering
	\includegraphics[width=0.45\textwidth]{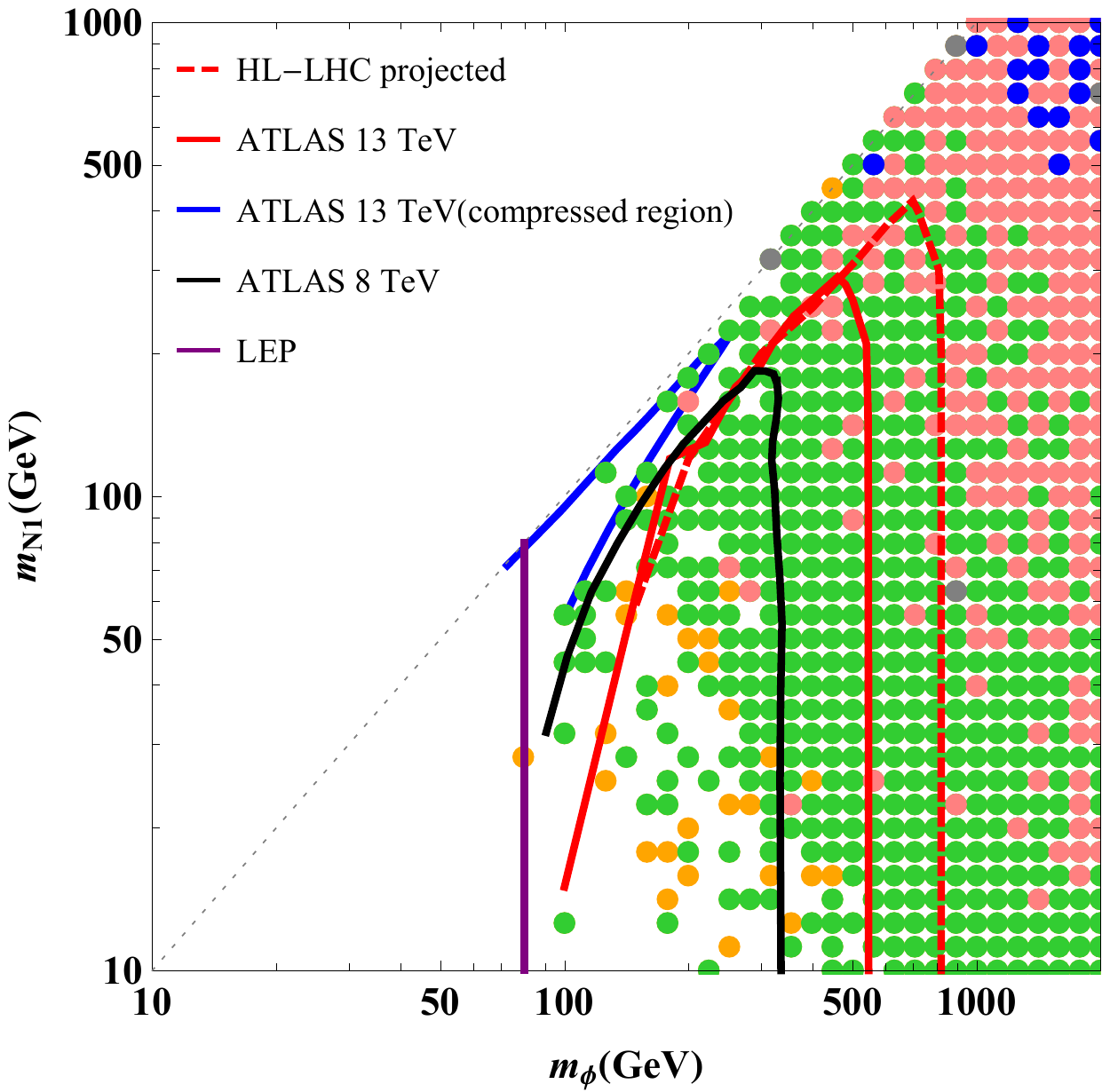}
	\caption{Masses distribution on the $m_{N1}-m_\phi$ plane, dots in orange, gray, pink and blue colors are excluded by LFV and dark matter detection experiments, the remaining green dots satisfy all these experimental constraints, see the context for more details. Also given are the lower limits from $8~\rm TeV$(black line) and $13~\rm TeV$(red solid line) ATLAS results, the red dashed line stands for a projected lower limit for HL-LHC, the blue line stands for limit from $13~\rm TeV$ compressed region, purple line represents the LEP limit, i.e. $m_\phi>80~\rm GeV$.}
	\label{fig:ms}
\end{figure}

The charged scalar could be Drell-Yan produced and then decay into a final signal of $\ell^+\ell^-+\slashed{E}_T$, the ATLAS collaboration~\cite{Aad:2019vnb,Aad:2019qnd} had searched such a signature under the supersymmetry framework, i.e. slepton pair produced and decaying into SM charged lepton plus neutralino, the limits have been given based on the final charged leptons to be left-handed, right-handed or the summation of both, we choose the first one since in our model only the left-handed charged leptons couple to $\phi^\pm$. For the study looking at compressed mass region \cite{Aad:2019qnd}, the kinematic variable $R_{\rm ISR}$, ratio of the $\slashed{E}_T$ to the transverse momentum of hadronic initial-state radiation(ISR), is defined to target at this region, light-flavor sleptons are constrained to have masses above 251 GeV for a mass splitting of 10 GeV, and constraints will extend down to mass splittings of 550 MeV at the LEP slepton limits. The exclusion limits had been showed in Figure \ref{fig:ms}, with red solid line stands for the $13~\rm TeV$ lower limit~\cite{Aad:2019vnb} while blue line stands for limit in compressed region at the same center-of-mass energy~\cite{Aad:2019qnd}. We also show the limits from lower energy searching, the black line stands for lower limit from $8~\rm TeV$ ATLAS result~\cite{Aad:2014vma} and the purple line represents the LEP limit~\cite{Abbiendi:2013hk}, i.e. $m_\phi$ needs to be larger than $80~\rm GeV$.

We could see that the ATLAS and LEP limits could almost exclude out all the low mass region, with the scalar mass $m_\phi$ being excluded to as large as $\sim 550~\rm{GeV}$, while the dark matter mass $m_{N1}$ being excluded to the largest value at about $300~\rm{GeV}$. Here we have also given out a simple projection, based on the $13~\rm{TeV}$ search, of the exclusion limit of the high-luminosity LHC(HL-LHC), the limit is showed as red dashed line. One could see the HL-LHC can exclude $m_\phi$ to $\sim 820~\rm{GeV}$ while $m_{N1}$ is excluded to largest value of $\sim 420~\rm{GeV}$.  We have also shown the constraints from LFV, dark matter relic density, direct and indirect detections. The dots in orange are those in the excluding capabilities of future LFV experiments, the gray and pink dots are those excluded by current and future direct detection experiments, dots in blue are those excluded by CTA indirect detection experiment, the green dots could satisfy all the LFV and dark matter detection constraints. As we have mentioned in the above discussions, the excluded parameters from direct and indirect detections are more concentrated in the large mass region, due to the mass enhancement from the dipole interaction in the direct detection processes, and large annihilation cross section for larger dark matter mass in indirect detection processes. The ATLAS and LEP exclusion limits, show out a complementary excluding capability when compared to that from the direct and indirect detections.

\section{Conclusion}\label{sec:con}
In this work, we have made a detailed phenomenology discussion on the scotogenic Dirac model, with different observable signatures. Firstly, the charged scalar $\phi^\pm$ and heavy Dirac fermion $N_i$ could mediate the LFV processes. The LFV constraints from various experiments, especially the decays $\mu\to e\gamma$ and $\mu\to 3e$, bound the relevant Yukawa i.e. $y_\Phi$, to be small and show a hierarchical structure with different flavors. The smallness of $y_\Phi$ will then make the dark matter annihilation, to reach the correct relic density, mainly through another Yukawa, i.e. $y_\chi$.
The visible signal of dark matter annihilation is annihilating into charged lepton final states, but the observational effect is mostly too small to be detected, only the future experiment CTA is able to detect part of the large dark matter mass region. After that, we have studied the dark matter direct detection in this model, with $N_1$ being selected as the dark matter candidate. The lowest order contribution to dark matter direct detection is at the one-loop level, and the contributions have classified into $y_\chi$-related and $y_\Phi$-related processes. For the former process, we find that the direct detections are too weak to be observed, even for the future experiments; while for the latter processes, two main contributions exist, i.e. vector operator $\mathcal{O}_{\rm VV}$ and magnetic dipole operator $\mathcal{O}_{\rm mag.}$. The dipole interaction could dominate the $y_\Phi$-related direct detection processes, due to the mass enhancement. We have projected the limits from direct detection experiments on the $y_\Phi-m_{N1}$ plane, and find that most of the parameters, which satisfy the LFV and relic density constraints, are under the current detecting capabilities, and could be further detected by the future direct detection experiments. Due to the mass enhancement, exclusion capabilities on the large mass region have not been weakened too much by the decreased detecting sensitivity. Finally we make a discussion on the observable signature on collider, the charged scalar $\phi^+$ could decay into charged leptons and the signature of $\ell^+\ell^-+\slashed{E}_T$ have been searched by ATLAS. The limits on the $m_{N1}-m_\phi$ have excluded a big portion of the parameter space, which shows out a complementary exclusion capability compared to direct and indirect detections.

\acknowledgments
The work of Zhi-Long Han is supported by the National Natural Science Foundation of China under Grant No. 11805081, Natural Science Foundation of Shandong Province under Grant No. ZR2019QA021.

\bibliographystyle{JHEP}
\bibliography{DSS}

\end{document}